\documentclass[a4paper,12pt]{article}
\usepackage{graphicx}
\usepackage{amssymb,amsmath}
\newcommand{\braket}[1]{\langle {#1} \rangle}
\newcommand{\ket}[1]{\vert {#1} \rangle}
\newcommand{\ba}{\begin{array}}
\newcommand{\ea}{\end{array}}
\newcommand{\sbar}{\langle s\rangle}
\newcommand{\Ac}{A}
\newcommand{\AL}{A_L}
\newcommand{\Ispin}{\stackrel{{\rm I-spin}}{\longrightarrow}}
\def\be{\begin{equation}}
\def\ee{\end{equation}}
\def\bc{\begin{center}}
\def\ec{\end{center}}
\def\bea{\begin{eqnarray}}
\def\eea{\end{eqnarray}}

\def\nn{\nonumber}
\def\K3pi{\ensuremath{K\rightarrow 3\pi\,}}
\def\Kp00{\ensuremath{K^{+}\rightarrow\pi^{+}\pi^{0}\pi^{0}\,}}
\def\Kppm{\ensuremath{K^{+}\rightarrow\pi^{+}\pi^{+}\pi^{-}\,}}

\def\Ke4{\ensuremath{K^{+}\rightarrow\pi^{+}\pi^{-}e^{+}\nu}\,}
\def\p00{\ensuremath{\pi^{0}\pi^{0}\,}}
\def\pp0{\ensuremath{\pi^{+}\pi^{0}\,}}
\def\pppm{\ensuremath{\pi^{+}\pi^{-}\,}}
\def\pppp{\ensuremath{\pi^{+}\pi^{+}\,}}
\def\pipi{\ensuremath{\pi\pi\,}}

\def\M{\ensuremath{\mathcal M}}
\def\mpp{\ensuremath{m_{\pi^{+}}}}
\def\mpo{\ensuremath{m_{\pi^{0}}}}

\def\ra{\ensuremath{\rightarrow\,}}
\def\S{\ensuremath{\mathbf S}}
\def\T{\ensuremath{\mathbf T}}
\def\R{\ensuremath{\mathbf R}}
\def\I{\ensuremath{\mathbf I}}
\def\uno{\ensuremath{\mathbf 1}}

\def\Re{{\rm Re\,}}
\def\Im{{\rm Im\,}}

\catcode`@=11
\def\marginnote#1{}
\newcount\hour
\newcount\minute
\newtoks\amorpm
\hour=\time\divide\hour by60
\minute=\time{\multiply\hour by60 \global\advance\minute by-\hour}
\edef\standardtime{{\ifnum\hour<12 \global\amorpm={am}%
        \else\global\amorpm={pm}\advance\hour by-12 \fi
        \ifnum\hour=0 \hour=12 \fi
        \number\hour:\ifnum\minute<10 0\fi\number\minute\the\amorpm}}
\edef\militarytime{\number\hour:\ifnum\minute<10 0\fi\number\minute}
\def\draftlabel#1{{\@bsphack\if@filesw {\let\thepage\relax
   \xdef\@gtempa{\write\@auxout{\string
      \newlabel{#1}{{\@currentlabel}{\thepage}}}}}\@gtempa
   \if@nobreak \ifvmode\nobreak\fi\fi\fi\@esphack}
        \gdef\@eqnlabel{#1}}
\def\@eqnlabel{}
\def\@vacuum{}
\def\draftmarginnote#1{\marginpar{\raggedright\scriptsize\tt#1}}
\def\draft{\oddsidemargin 0.0truein
        \def\@oddfoot{\sl preliminary draft \hfil
        \rm\thepage\hfil\sl\today\quad\militarytime}
        \let\@evenfoot\@oddfoot \overfullrule 3pt
        \let\label=\draftlabel
        \let\marginnote=\draftmarginnote
   \def\@eqnnum{(\theequation)\rlap{\kern\marginparsep\tt\@eqnlabel}%
\global\let\@eqnlabel\@vacuum}  }
\catcode`@=12
\begin{document}
\begin{titlepage}
\vspace*{-1cm}
\vskip 1.0cm
\begin{flushright}
January, 2005 \\
CERN-PH-TH/2005-012 
\end{flushright}
\vskip 3.0cm

\bc{\Large \bf Pion-pion scattering \\[0.2 cm] and the \K3pi decay amplitudes}

\ec
\vskip 1.0  cm
\begin{center}
{\large Nicola Cabibbo}${}^{a,b}$ and  {\large Gino Isidori}${}^{c}$ \\ [10 pt]
${}^{a}~$\textsl{CERN, Physics Department, CH-1211 Geneva 23, Switzerland} \\ [5 pt]
${}^{b}~$\textsl{Dipartimento di Fisica Universit\`a di Roma ``La Sapienza'' and \\ 
            INFN, Sezione di Roma, P.le A. Moro 2, I-00185 Rome, Italy} \\ [5 pt]
${}^{c}~$\textsl{INFN, Laboratori Nazionali di Frascati, I-00044 Frascati, Italy} \\ [5 pt]
\end{center}
\vskip 1.0cm
\begin{abstract}
\noindent
We revisit the recently proposed method for the determination of the $\pi\pi$ scattering 
length combination $a_{0}-a_{2}$, based on the study of the \p00 spectrum in \Kp00. 
In view of a precision measurement, we discuss here the effects due to smaller absorptive 
contributions to the \Kp00 and \Kppm amplitudes. We outline a method 
of analysis that can lead to a precision determination of $a_{0}-a_{2}$ and 
is based on very general properties of the S matrix.
The discussion of final-state rescattering and cusp effects is 
also extended to the two $K_L \to 3\pi$ coupled channels.
Thanks to the present work, the theoretical error on the 
$a_{0}-a_{2}$ combination extracted from the \Kp00 spectrum 
is reduced to about $5\%$. A further reduction requires the 
evaluation of the effects arising from radiative corrections.
\end{abstract}
\end{titlepage}

\setcounter{footnote}{0}
\vskip2truecm
\setlength{\baselineskip}{.7cm}


\section{Introduction}
In \cite{Cabibbo:2004gq} it was shown how  rescattering of the final state 
pions in \Kp00 produces a prominent cusp\footnote{~The existence 
of this cusp was first discussed in \cite{Meissner:1997fa}.}
in the total energy spectrum of the \p00 pair, 
whose amplitude is proportional to  the 
$a_{0}-a_{2}$ combination of the \pipi\  S-wave 
scattering lengths. 

The combination $a_{0}-a_{2}$ is a very interesting quantity: a benchmark 
observable to determine the structure of the QCD vacuum
and one of the few non-perturbative parameters which can be 
predicted with excellent accuracy from first principles.
Recent calculations~\cite{Colangelo}, 
that combine Chiral Perturbation Theory (CHPT)~\cite{ChPT} 
with Roy equations~\cite{Roy,Ananthanarayan}, 
lead to the precise prediction $(a_{0}- a_{2})\mpp = 0.265\pm0.004$.
So far, this high theoretical precision has not been matched 
by a similar experimental accuracy. The best 
direct information on \pipi\ scattering lengths
is the one extracted from $K_{e4}$ decays by
the BNL-E865 experiment~\cite{Pislak}, which 
is affected by a sizable ($\sim 6\%$) statistical error. 
Given the intrinsic statistical limitation of $K_{e4}$ decays 
with respect to the dominant $K\to 3\pi$ modes, and the 
different nature of systematical errors (including the 
theoretical ones) involved the extraction of 
\pipi\  scattering lengths in the two cases, 
it is definitely worth to explore  in more detail the 
proposal of Ref.~\cite{Cabibbo:2004gq}.

The NA48 runs of 2003 and 2004 have produced $\approx 10^{8}$ \Kp00\ decays, 
of which a few millions are in the \pppm threshold region with excellent $s_{\pipi}$ 
resolution. Since the cusp induced by the $\pi\pi$ rescattering 
is a $\approx 10\%$ effect on the \p00 spectrum, at a pure statistical level 
it should be possible to determine its amplitude with $\approx 1-2\%$  accuracy. 
In order to extract from this measurement a value for 
$a_{0}-a_{2}$ with a similar precision, it is necessary  
to reduce the theoretical uncertainties of the simple 
analysis proposed in \cite{Cabibbo:2004gq}.  
The present paper is a first step in this direction. 
 
As already noted in \cite{Cabibbo:2004gq}, 
the procedure presented there was incomplete for three main reasons:
\begin{enumerate}
\item It did not take into account the effect of radiative corrections.
\item It omitted higher terms in $v$, e.g.  $v^{3}$ terms in the imaginary part of the amplitude.
\item It omitted contributions from higher order rescattering effects.
\end{enumerate}
The effect of radiative corrections to the \Kp00 decay in general,  
and to the amplitude of the cusp in the \p00 spectrum, will not be discussed here. 
The radiative corrections to \K3pi decays have been evaluated (see e.g. Ref.~\cite{rad_cor}) 
to be of the few percent level, and dominated by Coulomb corrections. 
We expect radiative corrections to the cusp amplitude to be not larger than this. 

The second omission is minor, as the value of  $a_{0}-a_{2}$ is given 
by the term proportional to $v$, while the term in $v^{3}$ can be introduced 
as a free parameter in the experimental fit, and its possible prediction 
in CHPT is of lesser importance than that 
of the scattering lengths. The evaluation presented here includes these effects.

In this paper we will concentrate on correcting the third  omission. 
We will show how the unitarity and analyticity 
of the \S\  matrix elements can lead to a systematic expansion 
of the \Kp00 and \Kppm amplitudes in powers of the \pipi\  scattering lengths.  
The usefulness of this expansion 
derives from the relative weakness of \pipi\   scattering
that, in turn,  is a general consequence of the pseudo-Goldstone-boson nature of the pions 
and of the smallness of light-quark masses (or, in one word, a general consequence of CHPT). 
Rescattering effects in $K\to 3 \pi$ decays have already been widely discussed in the
literature in the framework of CHPT (see e.g. Ref.~\cite{K3p_CHPT,K3p_CHPT_IB}). However, 
previous analyses have been performed only up to next-to-leading order in the 
chiral expansion and -- with the exception of  Ref.~\cite{K3p_CHPT_IB} --
ignoring isospin breaking effects.
The approach presented in this paper is more general and particularly well 
suited to discuss the cusp amplitude: we shall use the effective field theory 
only for an explicit estimate of the irreducible $3\pi\to3\pi$ rescattering
(that turns out to be negligible). For completeness, we shall 
also present a general parameterization of rescattering effects 
and cusp amplitudes in $K_L \to 3\pi$ decays.

The paper is organized as follows: in section~\ref{sect:first} 
we shall introduce the definition of $\pi\pi$ scattering 
lengths used in the rest of this work, and we shall
recall some basic properties of the S matrix.
Section~\ref{sect:pipi} is devoted to 
analyse the consequences of unitarity and analyticity 
on the structure of various $\pi\pi\to\pi\pi$ amplitudes.
In section~\ref{sect:K3pi} we shall present the 
systematic expansion of  $K\to 3 \pi$ 
amplitudes in powers of the \pipi\  scattering lengths
up to $O(a_i^2)$, we shall also briefly discuss
possible strategies for the data analysis.
The results are summarized in the conclusions.

\section{\pipi\  Scattering}
\label{sect:first}
\subsection{Two pion states}
Consider the \S\ matrix element:
\begin{align}
	\label{Tfi}\braket{c,\vec p_{c};d,\vec p_{d}|\S |a, \vec p_{a};b,\vec  p_{b}} &=
	 \braket{c,\vec p_{c};d,\vec p_{d}|a, \vec p_{a};b, \vec p_{b}} + i \delta^{4}(P_{f}- P_{i}) 
	 \frac{\M_{fi}}{\sqrt{\prod{2E_{i}}}}
\intertext{The normalization of the states is chosen as}
	\label{norm1}\braket{c,\vec p_{c};d,\vec p_{d}|a, \vec p_{a};b, \vec 
		p_{b}}&=\delta_{ca}\,\delta_{db}\,\delta^{3}(\vec p_{c}-\vec p_{a})\, 
	\delta^{3}(\vec p_{d}-\vec p_{b}) +\delta_{cb}\,\delta_{da}\,\delta^{3}(\vec p_{c}-\vec p_{b})\, 
	\delta^{3}(\vec p_{d}-\vec p_{a})
\intertext{This normalization is compatible with the field theoretical definition:}
	\ket{a, \vec p_{a};b, \vec p_{b}}&=a^{\dagger}_{a}(\vec p_{a})a^{\dagger}_{b}(\vec p_{b})
		\ket{\Omega}\\
	[a_{a}(\vec p_{a}),\,a^{\dagger}_{b}(\vec p_{b})]&=\delta_{ab}\,\delta^{3}(\vec p_{b}-\vec p_{a})
\intertext{If we change variables to total and relative four-momentum,}
	P=p_{a}+p_{b};\quad k& =(p_{a}-p_{b})/2\quad 
		P'=p_{c}+p_{d};\quad  k'=(p_{c}-p_{d})/2\\
		\delta^{3}(\vec p_{c}-\vec p_{a})\, \delta^{3}(\vec p_{d}-\vec p_{b})&=
	\delta^{3}(\vec P'-\vec P)\, \delta^{3}(\vec k'-\vec k)\\
	\delta^{3}(\vec p_{c}-\vec p_{b})\, \delta^{3}(\vec p_{d}-\vec p_{a})&=
	\delta^{3}(\vec P'-\vec P)\, \delta^{3}(\vec k'+\vec k)\\
  \intertext{we can define the S-wave state in the center of mass, $\vec P = 0$}
	\label{Swave}\ket{\vec P, q, a,b}&=\frac{1}{\sqrt{4\pi}q} \int d^{3}\vec k \;\delta(q-|\vec k|)\;\ket{\vec P, \vec k, a,b}
\intertext{and verify that the normalization is}
	\label{norm2}\braket{\vec P', q', c, d|\vec P, q, a,b}&=\delta^{3}(\vec P'-\vec P)\delta(q'-q)
	\left( \delta_{ca}\delta_{db}+\delta_{cb}\delta_{da}\right)
\end{align}

\subsection{Isospin states}
We will adopt a phase convention, inspired by a field theoretical treatment, where
\begin{equation}
\I^{-}\ket{\pi^{+}}=-\sqrt{2}\ket{\pi^{0}};\quad \I^{-}\ket{\pi^{0}}=\sqrt{2}\ket{\pi^{-}}
\end{equation}
We note that this convention is different
from that used in early I-spin analysis of \K3pi decays, e.g. in ref. \cite{Weinberg:1960} and \cite{BartonKacser:1963}, but coincides with the one
adopted in CHPT studies of these decays, e.g. ref. \cite{Cronin:1967jq}.
For $I_{3}=0$ we then find the three states
\begin{eqnarray}
\ket{2,0}&=&\frac{\ket{\pi^{+}\pi^{-}}+\ket{\pi^{-}\pi^{+}}-2\ket{\pi^{0}\pi^{0}}}{\sqrt{6}}\\
\ket{1,0}&=&\frac{\ket{\pi^{+}\pi^{-}}-\ket{\pi^{-}\pi^{+}}}{\sqrt{2}} \quad\\
\ket{0,0}&=&\frac{\ket{\pi^{+}\pi^{-}}+\ket{\pi^{-}\pi^{+}}+\ket{\pi^{0}\pi^{0}}}{\sqrt{3}}
\end{eqnarray}
And for $I_{3}=1$,
\begin{eqnarray}
\ket{2,1}&=&\frac{\ket{\pi^{+}\pi^{0}}+\ket{\pi^{0}\pi^{+}}}{\sqrt{2}} \quad\\
\ket{1,1}&=&\frac{\ket{\pi^{+}\pi^{0}}-\ket{\pi^{0}\pi^{+}}}{\sqrt{2}} \quad
\end{eqnarray}
These states are normalized as (see eq.~\ref{norm1}, but note that the $\I=1$ states vanish for S-waves)
\be
	\braket{\vec P', q', I', I'_{3}|\vec P, q,I,I_{3}}= 
	2\delta_{I'I}\,\delta_{I'_{3}I_{3}}\,\delta^{3}(\vec P'-\vec P)\,\delta(q'-q)
\ee

\subsection{Low energy scattering and scattering lengths}
\label{sub:scatt_def}
S-wave scattering means that the $\M_{fi}$ defined in \eqref{Tfi} does not depend on the direction
of the relative momentum $\vec k$, but at most is a function of the CM energy $E$ or momentum $q$.
We than find easily that, working in the C.M. frame ($\vec P=0)$, 
\begin{align}
	\braket{\vec P', q', f|(\S-1)|\vec P, q,i}&=4\pi\, i q\,q' \delta(E'-E)\delta^{3}(\vec P'-\vec P)
	\frac{\M_{fi}}{\sqrt{\prod{2E_{i}}}}\\
	&=4\pi\, i \frac{q\,E'_{1}E'_{2}}{(E'_{1}+E'_{2})}\, \delta(q'-q^{*})\delta^{3}(\vec P'-\vec P)
	\frac{\M_{fi}}{\sqrt{\prod{2E_{i}}}}\\
	&=\pi\, i \frac{q\,}{(E'_{1}+E'_{2})}\, \delta(q'-q^{*})\delta^{3}(\vec P'-\vec P)
	\M_{fi}
\intertext{with $q^{*}$ the C.M. momentum required by energy conservation. In the non relativistic limit, }
	\label{nr1}\braket{\vec P', q', f|(\S-1)|\vec P, q,i}&=
	4\pi\, i q\,\mu\, \delta(q'-q^{*})\delta^{3}(\vec P'-\vec P)\frac{\M_{fi}}{\sqrt{\prod{2E_{i}}}}\\
	&=\pi\, \frac{i q}{2m_{\pi}} \delta(q'-q^{*})\delta^{3}(\vec P'-\vec P)\M_{fi}
\end{align}
Neglecting $\pi^{+}\pi^{0}$ mass differences, $\mu=m_{\pi^{+}}/2$.
For exact I-spin (that only makes sense in the limit of equal masses) we must have near threshold:
\be
	\S \ket{P,q,I,I_{3}}=\exp(2iqa_{I})  \ket{P,q,I,I_{3}}
\ee
so that
\be
	\label{nr2}\braket{\vec P', q', I', I'_{3}|(\S-1)|\vec P, q,I,I_{3}}\approx 
	4 i q a_{I}\,\delta_{I'I}\,\delta_{I'_{3}I_{3}}\,\delta^{3}(\vec P'-\vec P)\,\delta(q'-q)+ O(q^{2})
\ee
Comparing now \eqref{nr1} and \eqref{nr2} we find that near threshold
\be
	\M_{I\,I}\approx \frac{8a_{I}m_{\pi}}{\pi}
\ee
Identifying $m_{\pi}$ with $m_{\pi^{+}}$ we will thus \emph{define}: 
\begin{align}
%
%
	\label{a00}
 	\p00\to\p00 && \Re\M_{00} &= \frac{8a_{00} m_{\pi^{+}}}{\pi }&&
		(\pppm\ {\rm threshold}) &a_{00}& \Ispin \ \frac{a_{0}+2a_{2}}{3} \\
	\label{a0p}
 	\pi^{+}\pi^{0}\to\pi^{+}\pi^{0} && \Re\M_{+0} &= \frac{8a_{+0} m_{\pi^{+}}}{\pi }&&
		(\pp0\ {\rm threshold}) &a_{+0}& \Ispin \ \frac{a_{2}}{2}\\
	\label{a02}
 	\pi^{+}\pi^{-}\to\pi^{0}\pi^{0} && \Re\M_{x} &= \frac{8a_{x} m_{\pi^{+}}}{\pi }&&
		(\pppm\ {\rm threshold} &a_{x}&  \Ispin \ \frac{a_{0}- a_{2}}{3} \\
	\label{apm}
 	\pi^{+}\pi^{-}\to\pi^{+}\pi^{-} && \Re\M_{+-} &= \frac{8a_{+-} m_{\pi^{+}}}{\pi }&&
		(\pppm\ {\rm threshold}) &a_{+-}&\Ispin \ \frac{2a_{0}+a_{2}}{6}\\
	\label{app}
 	\pi^{+}\pi^{+}\to\pi^{+}\pi^{+} && \Re\M_{++} &= \frac{8a_{++} m_{\pi^{+}}}{\pi }&&
		(\pppp\ {\rm threshold}) &a_{++}& \Ispin \ a_{2}
\end{align}
For each process we have noted the threshold at which the scattering length is defined,
and the value it would have in the limit of exact I-spin symmetry.
  
The problem that must at some time be faced in comparing the result of cusp studies 
to the CHPT prediction for $a_{0}-a_{2}$ is that of taking into account radiative 
corrections. Note that the threshold region is one where I-spin is maximally broken. 
We will take the point of view that the quantity $a_x$ introduced in eq.~\eqref{a02} should  
be taken as a definition of the effective scattering-length combination 
measured from the cusp effect.
The experimentally determined value for this quantity should be compared with a 
CHPT prediction which \emph{includes} the effects of radiative corrections 
and I-spin breaking due to $m_u\not=m_d$. The evaluation of these 
subleading effects can be subdivided into two separate tasks: computing their
impact on the CHPT predictions of the various $\pi\pi \to \pi\pi$ 
amplitudes in eqs.~\eqref{a00}--\eqref{app}, and determining how 
they would affect the decomposition of the  $K\to 3 \pi$ amplitude 
presented in this paper.
The first point has already been partially addressed in the 
literature \cite{Urech,Gall} -- it turns out to be only a few percent correction 
in the case of $a_x$ \cite{Gall} -- and can be easily implemented in our 
decompostion. However, at the moment we are lacking of
a consistent description of the second point, or the 
evaluation of I-spin breaking effects in the relation between  $\pi\pi \to \pi\pi$ 
and $K\to 3 \pi$ amplitudes.

In the following we will proceed using eqs.~\eqref{a00}--\eqref{app} 
as a definition of the different scattering length \emph{combinations}.
We shall use their expressions in terms of $a_{0}, a_{2}$ only as a first 
approximation, pending a consistent evaluation of all the I-spin breaking effects. 
Note the use of $m_{\pi^{+}}$ in these definitions, and of the \pppm threshold\footnote{~At 
this threshold the cusp correction to the $\p00\to\p00$ 
scattering amplitude vanishes. } in \eqref{a00}. 

\medskip

In the case where the scattering occurs well above threshold, 
eqs. \eqref{a00}--\eqref{app} should be modified to take into 
account the non trivial kinematical dependence of $\M_{fi}$. 
Expanding up to linear terms   
in the kinematical variables $s$, $t$ and $u$, we can neglect all 
higher modes but the P wave. The generic matrix element takes the 
form\footnote{~This expression does not include the effects of 
threshold singularities, whose structure will be discussed in section~\ref{sect:pipi}.}   
\be
\Re \M_{ij} = \frac{ 8 m_{\pi^{+}}  a_{ij} (s)  }{\pi } + \frac{6  m_{\pi^{+}} a^P_{ij} }{  \pi}~ 
 \frac{  (t-u)}{ m^2_{\pi^{+}}}   
\ee
with
\be
a_{ij} (s) = a_{ij} \left[ 1+ r_{ij} \frac{(s- s_{\rm threshold})}{ 4 m^2_{\pi^{+}}} \right]
\ee  
The $a_{ij}$ are the combinations of constant S-wave scattering lengths
defined in eqs. \eqref{a00}--\eqref{app}, while the $r_{ij}$ 
define the corresponding effective ranges. In the isospin limit, we can express 
the $r_{ij}$ in terms of the effective ranges of $a_0$ and $a_2$,
following the isospin decomposition reported in  eqs. \eqref{a00}--\eqref{app}.
According to the detailed analysis of Ref.~\cite{Colangelo}, 
these are given by $r_{0}=1.25 \pm 0.04$  and $r_{2}=1.81\pm 0.05$,
values which are consistent with those recently reported in 
Ref.~\cite{Ind} and also not too far from the lowest-order 
CHPT predictions $r^{(2)}_{0}=8/7$ and $r^{(2)}_{2}=2$.

The only two channels with non vanishing $a^P_{ij}$ are 
the $\pi^{+}\pi^{-}\to\pi^{+}\pi^{-}$ and $\pi^{+}\pi^{0}\to\pi^{+}\pi^{0}$ ones.
In the I-spin limit
\be
a^P_{+-} = a^P_{+0} = a_1/2~,
\ee
while the lowest-order CHPT prediction is  
$a^{(2)}_1 = m_\pi^2/(12 \pi f_\pi^2)$.

\subsection{Cluster decomposition and the Operator notation.}
The \S-matrix elements can in general be cluster-decomposed\footnote{~For 
a discussion of cluster decomposition, see e.g.  \cite{WichmanCrichton63,Taylor66}, 
and \cite[Vol~I,~Ch.~3] {WeinbergQFT}.}\ %
 into the sum of a connected part 
(in perturbation theory this is the sum of the connected diagrams),
and one or more terms that are the product of connected terms, and correspond to the separate interaction of non overlapping subsets of the  initial particles to yield  non overlapping subsets of the final particles. Among the disconnected terms there may be some where one or more  of the initial particles propagate without interacting at all.

It will be convenient to express the \S\, and \T\, operators
in terms of creation and annihilation operators for asymptotic states, so that
 we can write \T\, as a sum of operators:
\be\S = \uno+i\T;\quad\quad\quad \T = \sum_{m,n}\T_{m,n} \ee
Each of these operators can be expressed as:
\be 
	\T_{m,n}= \frac{1}{m!\,n!}\sum_{f\,i}  \int \prod \left[ d^{3}p\right] \delta^{4}(P_{f}-P_{i})
	a^{+}_{f_{1}}\ldots a^{+}_{f_{m}}a_{i_{1}}\ldots a_{i_{n}} \frac{\M_{fi}}{\sqrt{\prod{2E_{i}}}}
\ee
where the sum is over particle types and the integral is over the
three-momenta of the initial and final particles.
We note that each $\T_{m,n}$ can contribute to an $n\to m$ transition, but also to  an $n+k\to m+k$  transition, 
with $k$ particles passing through without interacting with the others. 
In general the $\T_{m,n}$ term will contain a connected part, $\T^{C}_{m,n}$, where the corresponding $T_{fi}$ 
contains a single $\delta^{4}(P_{f}-P_{i})$ factor, and other terms with two or more such momentum
 conservation factors, and can be expressed as well ordered products (annihilation operators on the right) 
of  ``smaller''  $\T^{C}_{m',n'}$. For instance in the case $m=n= 4$ we would find
 \be i\T_{4,4} =i\T^{C}_{4,4}+:\left(i\T_{2,2}\right)^{2}: \ee
In the case we will be interested in, the \Kp00 and \Kppm decays, we will be working with 
$\T_{m,n}$ terms that coincide with their connected parts. 
We do not thus need to explore the disconnected parts of  $\T_{m,n}$ in more detail here.


\subsection{Time reversal symmetry}

We will neglect the effects of time reversal and $CP$ violation on $K\to 3\pi$ decays.  We have
very strong experimental limits on these effects, and the theoretical expectation is even smaller.
Time reversal symmetry  implies the relation
\be \braket{B|\S|A} = \braket{A_{T}|\S|B_{T}}                               \ee
where $\ket{B_{T}},\ket{A_{T}}$ are the ``time reversed states'', that for pseudoscalar mesons 
amounts to changing the sign of all momenta, $\vec p\to-\vec p$. 
In the case of \K3pi and \pipi$\to$\pipi,   
that arise in the following discussion, we can change the sign of all momenta with a combination 
of Lorentz transformations and rotations, so that  we have simply
\be 
	\label{Trev}\braket{B|\S|A} = \braket{A|\S|B} \quad\quad\quad
\ee
Because of parity conservation, the same condition holds for the 
$3 \pi \to 3 \pi$ strong re-scattering. 
So that, neglecting $CP$ breaking effects, \S~ is symmetric
for all the cases of interest for this analysis.


\section{Unitarity, analyticity and the \pppm threshold}
\label{sect:pipi}

Due to the presence of the square-root singularity connected to the \pppm threshold within 
the phase space for 
\Kp00, we have to distinguish the two zones above and below the  \pppm threshold. 
We can write the amplitude $\M_{fi}$ above the threshold in the form: 
\begin{align}
    \M_{fi} &= A + B\sqrt{\frac{s_{3}-4m_{\pi^{+}}^{2}}{s_{3}}} && s_{3}>4m_{\pi^{+}}^{2}
    \label{TAbove}\\
    \intertext{where both $A$ and $B$ are regular at the \pppm threshold. 
    This expression can be analytically continued below the threshold, where it becomes}
  \M_{fi} &= A + iB \sqrt{\frac{4m_{\pi^{+}-s_{3}}^{2}}{s_{3}}} && s_{3}<4m_{\pi^{+}}^{2}
    \label{TBelow}
\end{align}  

Applying unitarity above the threshold we can determine the imaginary parts of both $A$ and $B$.
Also,  unitarity below the threshold determines the \emph{real part} of $B$. The experimental data can then be analysed with the following procedure:
\begin{enumerate}
\item Parametrize the real part of $A$ as a polynomial in the three independent kinematical variables,
$s_{1}, s_{2}, s_{3}$, as outlined in \cite{PDBook}.
\item Parametrize  the \pipi\  scattering amplitudes  in terms of the  scattering lengths and possibly additional parameters.
\item Use unitarity to derive $B$ and the imaginary part of $A$. 
\end{enumerate}

It is best to work out the consequences of  unitary in the operator formalism, where we can express the \S\ operator in terms of the hermitian and anti-hermitian parts of the $\T$ operator,
\begin{align} \S &= \uno+i(\R+i\I) 
     &\text{then unitarity implies}\\
	2 \I &= \R^{2}+\I^{2}& \text {or, solving for } \I,\\
	\I &= \uno -\sqrt{\uno-\R^{2}} &\text {and as a power series in } \R,\\
	\label{R2I}
	 \I &= \frac{1}{2} \R^{2} + \frac{1}{8} \R^{4} +\frac{1}{16} \R^{6} +\frac{5}{128} \R^{8} ...
\end{align}
Time reversal invariance implies that \S~is symmetric, so that the matrix elements of \R~and 
\I~correspond directly to the real and imaginary parts of \T~matrix elements.

The last equation allows for a systematic computation of the imaginary parts in terms of the real parts of the scattering amplitudes. The utility of this expansion derives from the assumed smallness of the \pipi\  scattering lengths. In the case of \Kp00 the first term in the development yields terms  $\propto a_{i}$ and
$\propto a_{i}^{2}$, and higher, while the further terms, 
starting with $\R^{4}$, will contribute  corrections 
$\propto a_{i}^{3}$ and higher.   


\subsection{\p00 scattering}
Let us apply the ideas outlined above to \p00 scattering. We will work in the threshold region, so that we can neglect higher partial waves and any dependence of the amplitude $\M_{00}$ on the $t,u$ variables. We will also neglect higher (e.g. $4\pi$) cuts, and only retain the first term in eq. \eqref{R2I}, so that we will neglect terms $O(a_{i}^{4})$ and higher. We  will also use $m_{\pi^{+}}$ as unit of energy, so that we write e.g. $a_{00}$ instead of  $a_{00}m_{\pi^{+}}$. 
 
Let us start by defining the ``velocities'',
\begin{align}
	v_{\pm}(s)&=\sqrt{\frac{|s-4m_{\pi^{+}}^{2}|}{s}}\\
	v_{00}(s)&=\sqrt{\frac{|s-4m_{\pi^{0}}^{2}|}{s}}
\end{align}
We can then write, in analogy to eqs. \eqref{TAbove}, \eqref{TBelow},
\begin{align}
    \M_{00} &= A_{00} + B_{00}v_{\pm}(s)&& s>4m_{\pi^{+}}^{2}
    \label{PiAbove}\\
  \M_{00} &= A_{00} + iB_{00} v_{\pm}(s) && s<4m_{\pi^{+}}^{2}
    \label{PiBelow},
    \intertext{and, for $\pi^{+}\pi^{-}\to \pi^{0}\pi^{0}$,}
    \M_{x} &= A_{x} + B_{x}v_{\pm}(s)&& s>4m_{\pi^{+}}^{2}    
\end{align}  
 where 
$A_{00},B_{00},A_{x},B_{x}$ are regular at the \pppm threshold. We can express $\Re (A)$ as a polynomial in $s$. We can simply write:
 \begin{align} 
	\Re(A_{00}) &= \frac{8a_{00}(s)}{\pi }\,;
	&&\text{where }a_{00}(4m_{\pi^{+}}^{2})=a_{00}\\
	\intertext{and similarly for $\pi^{+}\pi^{-}\to \pi^{0}\pi^{0}$,}
	\label{Mx}
	\Re(A_{x}) &= \frac{8a_{x}(s)}{\pi }\,;
	&&\text{where }a_{x}(4m_{\pi^{+}}^{2})=a_{x}
\end{align} 
The \pppm intermediate state contributes to $\Im\M_{00}$ only above the \pppm threshold,
while the \p00 state contributes both above and below, so that we find 
\be \label{ab}
	\Im\M_{00}= \frac{\pi}{4}v_{\pm}(s)\left(\Re\M_{x}\right)^{2}\Theta(s-4m_{\pi^{+}}^{2})
		+ \frac{\pi}{8}v_{00}(s)\left(\Re\M_{00}\right)^{2} +O(\R^{4})
\ee	
where $O(\R^{4})$ indicates the neglected higher terms in eq.  \eqref{R2I}.  Evaluating eq. \eqref{ab} above the \pppm threshold, and neglecting terms $O(\R^{4})$,  this translates into
\begin{align}
   \label{IA00-1}\Im A_{00}&=\frac{\pi v_{00}}{8}[(\Re A_{00})^{2}
   	+\frac{s-4m_{\pi^{+}}^{2}}{s}(\Re B_{00})^{2}]
   			+\frac{ \pi}{2} 
                 \Re A_{x}\Re B_{x}\frac{s-4m_{\pi^{+}}^{2}}{s}\\
   \label{IB00}\Im B_{00}&=\frac{\pi }{4}[(\Re A_{x})^{2}+
   	\frac{s-4m_{\pi^{+}}^{2}}{s}(\Re B_{x})^{2}
  			+v_{00}(s)\Re A_{00}\Re B_{00}] 
  \intertext{and evaluating it below the \pppm threshold}
  \label{IA00} \Im A_{00}&=\frac{\pi v_{00}(s)}{8}[(\Re A_{00})^{2}+
  	\frac{4m_{\pi^{+}}^{2}-s}{s}(\Im B_{00})^{2}] 	
  	\\
 \label{RB00} \Re B_{00}&=-\frac{\pi v_{00}(s)}{4}\Re A_{00}\Im B_{00} 
\end{align}
	From eqs. \eqref{IB00},\eqref{RB00} we see that $\Im B_{00}=O(\R^{2})$ and 
$\Re B_{00}=O(\R^{3})$, so that, comparing eqs. \eqref{IA00} and \eqref{IA00-1} we conclude that also
	$\Re B_{x}$ is at least $O(\R^{3})$. Neglecting terms of $O(\R^{4})$,
the final result is
\begin{align}
	\label{IB00F}
	&\Im B_{00}=\frac{\pi }{4}(\Re A_{x})^{2}
		= \frac{16 }{\pi} \left(a_{x}(s)\right)^{2}\\
	\label{RB00-F} 
	&\Re B_{00}\;=
		-\frac{\pi ^{2}v_{00}(s)}{16}\Re A_{00}(\Re A_{x})^{2}
		= -\frac{32 v_{00}(s)}{\pi} a_{00}(s)(a_{x}(s))^{2}\\
	\label{IA00-F}& \Im A_{00}=\frac{\pi v_{00}(s)}{8}(\Re A_{00})^{2}
		= \frac{8 v_{00}(s) }{\pi} \left(a_{00}(s)\right)^{2}\\
	&\Re B_{x}\;\;\;= O(\R^{3})	\label{Bx}
\end{align}

\subsection{\pp0 and \pppp scattering}
In the following we will also need expressions for \pp0 and \pppp
scattering. The situation here is simpler, since at $O(\R^{2})$ there is  only one 
intermediate state. However, in this case we are interested in kinematical 
configurations where the amplitudes are not close to threshold and P-wave
contributions cannot be completely neglected.  Expressing the latter in terms 
of the I=1 scattering length in units of $m_{\pi^{+}}$ (we can safely 
neglect I-breaking corrections in this case), 
the real part of the amplitudes can be parametrized as
\bea
\label{Mp0}\Re(\M_{+0}) &=& \frac{8a_{+0}(s)}{\pi } +
   \frac{3 a_1}{\pi} \frac{ (t-u) }{m^2_{\pi^+} }   \,,		\\
\label{Mpppp}\Re(\M_{++}) &=& \frac{8a_{++}(s)}{\pi } \,,
\eea
We have adopted the standard notation 
$t=(p_1-p_1^\prime)^2$ and $u=(p_1-p_2^\prime)^2$, where 
$\pi^{+}(p_{1})\pi^{0}(p_2) \to \pi^+(p_1^\prime)+\pi^{0}(p_2^\prime)$.
The imaginary parts can easily be computed, but are not needed in the following.

\subsection{\pppm scattering}\label{sect:masscont}
In this section we will consider \pppm\ra\pppm scattering, that enters in rescattering corrections to
the \Kppm  decay amplitude. We will again work close to the threshold region, neglecting 
higher partial waves and the kinematical dependence from $t$ and $u$ variables.
Here we meet with a new problem: in the case of \p00\ra\p00 scattering we were able to apply unitarity  below the \pppm threshold, and this was used to derive a value for $\Re B_{00}$, eq. \eqref{RB00}, \eqref{RB00-F}. In the present case moving below the \pppm threshold implies an analytic continuation to an unphysical region. We will proceed to do this by considering a continuation in the $\pi^{+}$ and $\pi^{0}$ masses, a procedure that is certainly  legitimate in a field theory,
such as CHPT, where the $\pi^{+}-\pi^{0}$ mass difference can be changed by introducing an extra mass term in the Lagrangian. We will then work out the consequences of 
unitarity in a situation where $m_{\pi^{0}}>m_{\pi^{+}}$, and analytically continue the results to the situation where the masses have their physical value.

Assuming now that  $m_{\pi^{0}}>m_{\pi^{+}}$ we must distinguish the case where $s>4m_{\pi^{0}}^{2}$, where both \pppm and \p00 can appear as intermediate states, and 
that where $s<4m_{\pi^{0}}^{2}$ where only the \pppm state can contribute. We can write
\begin{align}
    \M_{+-} &= C_{\pm} + D_{\pm}v_{00}(s)&& s>4m_{\pi^{0}}^{2},
    \label{PMAbove}\\
  \M_{+-} &= C_{\pm} + iD_{\pm} v_{00}(s) && s<4m_{\pi^{0}}^{2}
    \label{PMBelow},
\end{align}  
As in the case of the \p00 scattering we start by defining the real part of $C_{\pm}$ in terms of the scattering length,
\be
	\Re(A_{\pm}) = \frac{8a_{\pm}(s)}{\pi }\,;
\ee
and applying unitarity at $O(\R^{2})$ we obtain:
\be \label{abPM}
	\Im\M_{+-}= \frac{\pi}{8}v_{00}(s)
	\left(\Re\M_{x}\right)^{2}\Theta(s-4m_{\pi^{0}}^{2})
		+ \frac{\pi}{4}v_{\pm}(s)\left(\Re\M_{+-}\right)^{2} +O(\R^{4})
\ee	
 Evaluating eq. \eqref{abPM} both above and below the \p00 threshold, and neglecting terms $O(\R^{4})$,  this translates\footnote{~We omit the intermediate steps that follow the lines of the preceding section.}
  into
\begin{align}
	\label{IDpmF}
	&\Im D_{\pm}=\frac{\pi }{8}(\Re A_{x})^{2}
		= \frac{8 }{\pi} \left(a_{x}(s)\right)^{2}\\
	\label{RDpmF} 
	&\Re D_{\pm}\;=
		-\frac{\pi ^{2}v_{\pm}(s)}{16}\Re A_{00}(\Re A_{x})^{2}
		= -\frac{32 v_{\pm}(s)}{\pi} a_{+-}(s)(a_{x}(s))^{2}\\
	\label{pm00F}& \Im C_{\pm}=\frac{\pi v_{\pm}(s)}{4}(\Re C_{\pm})^{2}
		= \frac{16 v_{\pm}(s) }{\pi} \left(a_{+-}(s)\right)^{2}
\end{align}
The continuation to the physical values of the $\pi^{+}$ and $\pi^{-}$ masses is simply achieved
by using in the above expressions the correct masses for $v_{\pm}, v_{00}$ and the physical 
values for the scattering lengths.

\section{$K \to 3 \pi$ decays}
\label{sect:K3pi}

In the following we shall apply the results of the previous section 
to describe rescattering effects in $\K3pi$ decays. 
Our main interest will be on the \Kp00 channel, where the cusp effect is most prominent 
and useful for the determination of $(a_{0}-a_{2})$, but we shall also consider the \Kppm
decay, whose amplitude is needed for the \Kp00 analysis. 
Similarly, we shall discuss the $K_L \to 3\pi^0$ decay,
where smaller cusp effects  --- still proportional to $(a_{0}-a_{2})$
and related to the $K_L \to \pi^+\pi^-\pi^0 \to 3\pi^0 $ process --- should also be visible.
As in the previous section, we shall consider rescattering effects 
only up to $O(a_{i}^{2})$ corrections to the leading amplitudes. 
To be more precise, we shall evaluate the full imaginary parts of the 
amplitudes at  $O(a_{i})$ and the corresponding $O(a^2_{i})$ corrections 
to the real parts. 

Similarly to the $\pi^0\pi^0\to\pi^0\pi^0$ case, we can decompose \Kp00 and \Kppm
amplitudes into a regular term and one that is singular at the \pppm threshold. 
We will use the standard kinematical variables $s_{i}=(p_K-p_{\pi_i})^2$, $i=1\ldots 3$,
as specified in the Particle Data Group Review \cite{PDBook}, 
the index ``3'' referring to the odd pion ($\pi^{+}$ or $\pi^{-}$ for the two decays). 
In particular, for \Kp00, $s_{3}$ coincides with the square of the CM energy 
of the \p00 pair. We will thus write
\begin{align}
    \M_{00+} &= A_{00+} + B_{00+}v_{\pm}(s_{3})&& s_{3}>4m_{\pi^{+}}^{2}~,
    \label{KAbove}\\
  \M_{00+} &= A_{00+} + iB_{00+}  v_{\pm}(s_{3}) && s_{3}<4m_{\pi^{+}}^{2}~.
    \label{KBelow}
    \intertext{For the \Kppm amplitude it will be convenient to separate the terms which 
contain the singularity in $s_{1}, s_{2}$ associated with the \p00 threshold, and write:}
    \label{Kppm}
    \M_{++-} &= C_{++-} + D^{(1)}_{++-}v_{00}(s_{1})+ D^{(2)}_{++-}v_{00}(s_{2})~,
\end{align}  
where  Bose symmetry implies that
\begin{align}  
	\M_{(i)}(s_{1},s_{2},s_{3})&= \M_{(i)}(s_{2},s_{1},s_{3}),
		&&(i=++-, 00+) \text{, that translates into} \\
	D_{++-}^{(1)}(s_{1},s_{2},s_{3})&= D_{++-}^{(2)}(s_{2},s_{1},s_{3}), &&\text{etc.}
\end{align}
$A_{00+}$, $B_{00+}, C_{++-},$ and $D_{++-}$ are expected to be analytic 
functions of  $s_{1},s_{2},s_{3}$ in the physical region for the two decays, 
with square-root singularities at the borders associated with different \pipi thresholds. 

Unitarity will allow us to express $\M_{00+}$ and $\M_{00+}$ in terms of $\Re A_{00+},\, \Re C_{++-}$
and $\pi\pi$ scattering lengths. In the case of  $\Re A_{00+},\, \Re C_{++-}$
we adopt a parametrization inspired by the PDG tables, namely
 \begin{align}
 	\Re A_{00+}(s_{1},s_{2},s_{3}) &=R^{0}(s_{3}) 
	= \Ac^{0} \left[ 1+g^{0}(s_{3}-s^0_{0})/2\mpp^{2}
		+\tilde h^{0}(s_{3}-s^0_{0})^{2}/2\mpp^{4} \right] \label{Mzero}\\
 	\Re C_{++-}(s_{1},s_{2},s_{3}) &= R^{+}\!(s_{3}) 
	= \Ac^{+} \left[ 1+g^{+}\!(s_{3}-s^+_{0})/2\mpp^{2}
		+\tilde h^{+}\!(s_{3}-s^+_{0})^{2}/2\mpp^{4} \right] \label{Mplus}	
\end{align}
where 
\be 
\sum_{i=1\ldots3} s_i = \left\{ \ba{ll} 
3 s_0^0  = m_K^2 + 2 m^2_{\pi^0} + m^2_{\pi^+} \quad &(\Kp00) \\
3 s_0^+ = m_K^2 + 3 m^2_{\pi^+} &(\Kppm) \ea \right.
\ee
The PDG tables also include terms proportional to $(s_{1}-s_{2})^{2}$, but their coefficients are small and compatible with zero\footnote{~The coefficient of the $(s_{1}- s_{2})^{2}$ term in $|\M_{00+}|^{2}$ is known to be very small: it is listed in the PDG tables \cite{PDBook} as $k=0.004\pm0.007$.
The $k$ coefficient for \Kppm is also compatible with zero, but with a slightly larger error.}.  They can be reintroduced, if needed for fitting a precise data-set, and one could also introduce higher powers of $(s_{3}-s_{0})$. We can compare this parametrization with that adopted in the PDG if we neglect the other contributions to the decay amplitude discussed in this paper, namely the imaginary parts of $A_{00+},C_{++-}$ and the whole of $B_{00+},D^{1,2}_{++-}$, that give smaller contributions to the decay rates, except in the cusp region of \Kp00. We then obtain:
\begin{align}
	g^{0,+} &\approx g_{\text{\tiny PDG}}; 
		& \tilde h^{0,+} &\approx h_{\text{\tiny PDG}}-( g^{0,+}/2 )^{2},
	\intertext{and, using the PDG average values,}
	g^{0} &\approx 0.638\pm 0.020; &h^{0}&\approx -0.051 \pm 0.01\\
	g^{+} &\approx -0.2154\pm 0.0065; &h^{+}&\approx 0. 0004\pm 0.004
\end{align}
Interestingly, the PDG values suggest a quadratic term that is vanishing small in the \Kppm amplitude. The small and negative value in the \Kp00 amplitude  could simply arise from the effect on the previous fits of an undetected cusp in that decay.  In this situation it would appear that the quadratic terms in eq. \eqref{Mzero} could be dropped at least in a first analysis that takes  into account the cusp effect and other absorptive contributions.

\subsection{Three-pion scattering and two-pion cuts}
\begin{figure}[t]
	\begin{center}
	\includegraphics[scale=0.85]{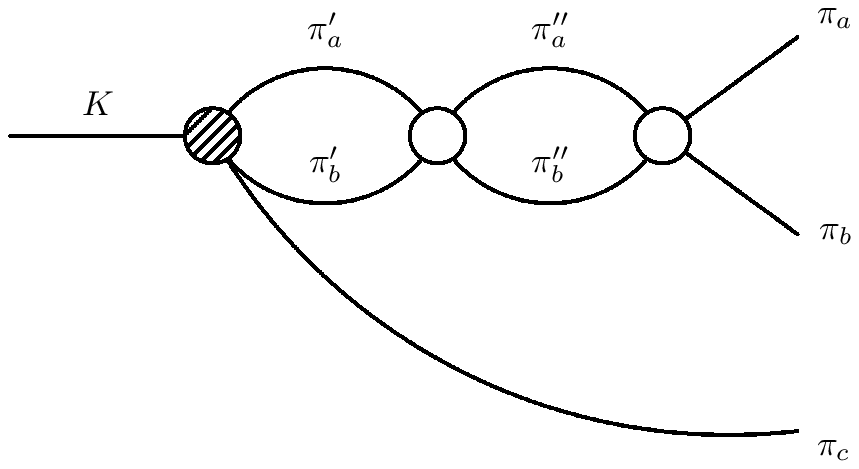}  \\
        \vspace{-1.5 cm} \hspace{-5.5 cm} a) \vspace{1.0 cm} \\
	\hspace{-0.5 cm}
        \includegraphics[scale=0.85]{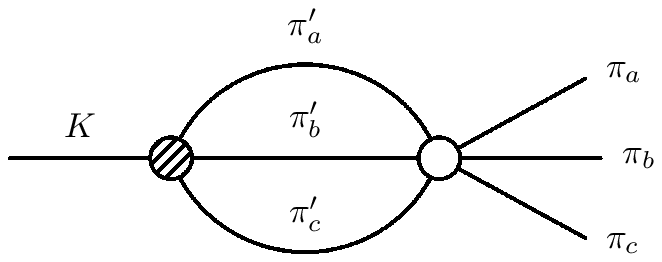}  
        \hspace{-1.3 cm}
	\includegraphics[scale=0.85]{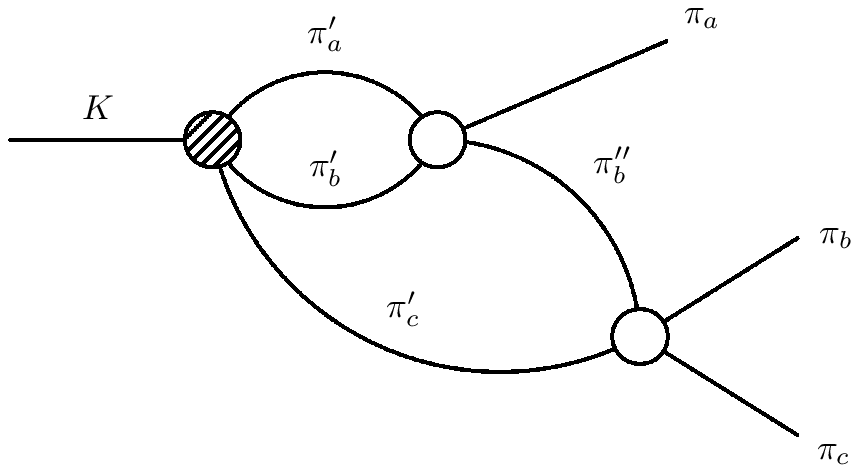}  \\ 
        \hspace{-2.0 cm} b) \hspace{7.0 cm}  c) \\
	\end{center}
	\caption{Examples of $K\to3\pi$ rescattering topologies at the two-loop level: 
a) single-channel $\pi\pi$ scattering; b) irreducible $3\pi\to3\pi$ contributions; 
c)  $3\pi\to3\pi$ amplitude due to multi-channel $\pi\pi$  scattering. \label{fig:3p} }
\end{figure}

Our final goal is the evaluation of rescattering effects -- and particularly 
the determination of the cusp amplitude -- in the three-pion states 
produced by $K$ decays. In general, in the case of $3\pi$ states, we can 
distinguish two basic contributions to the unitarity relations generated by \eqref{R2I}:
those arising from rescattering of a pair of pions 
in a given channel -- with a third spectator pion -- (see e.g. fig.~\ref{fig:3p}a)
and those due to $3\pi\to3\pi$ connected diagrams. 
At the level of approximation we are working, it is also convenient to distinguish 
between $3\pi\to3\pi$ one-particle-irreducible diagrams (fig.~\ref{fig:3p}b)
and  $3\pi\to3\pi$ reducible amplitudes due to multiple $\pi\pi$ 
scattering in different channels (fig.~\ref{fig:3p}c).

The $3\pi\to3\pi$ irreducible contribution is the only one that cannot be expressed in terms of 
$\pi\pi$ scattering lengths, but it turns out to be safely negligible. 
A simple and reliable estimate of its size can be obtained using the lowest-order 
CHPT Lagrangian to evaluate the $3\pi\to3\pi$ irreducible amplitude, and employing the 
non-relativistic approximation for the $3\pi$ states. 
In this limit, the irreducible scattering leads to a constant 
imaginary part of $O(10^{-4})$. For instance in the \Kp00 case we find 
\be
	\label{3to3}
	(\Im A_{00+})_{3\pi}  = -\frac{Q^{2}m_{\pi^{+}}^{2}}{360\pi^{2}f_{\pi}^{4}} 
        \langle \Re A_{00+} \rangle
	\approx  -4\times 10^{-4}\, \langle \Re A_{00+} \rangle
\ee
where $Q$ is the $Q$-value of the \Kp00 decay, $f_{\pi}=130.7 $ MeV is the pion decay constant,
and $\langle \Re A_{00+} \rangle$ is the average of the real part of the amplitude 
over the Dalitz plot.  This contribution, and others of a similar size also in the
other channels, appears to be safely negligible at  the $O(10^{-3})$ accuracy
for the decay rates we are aiming for.

The evaluation of the single-channel $\pi\pi$ scattering in $K\to 3 \pi$
decays proceeds exactly as for the $\M_{\pi\pi}$ amplitudes discussed 
in the previous section. To this purpose, it is useful to observe that 
all the previous results can be recovered in a diagrammatic framework 
by considering the absorptive two-pion cuts of appropriate Feynman diagrams. 
In particular, the  $O(\R^{3})$ contributions to the real part of the 
cusp amplitude (such as the expressions for $\Re B_{00}$ and 
$\Re D_\pm$ discussed before) can be derived by considering the
two s-channel cuts of diagrams similar to the one in fig.~\ref{fig:3p}a
(i.e.~setting both the $\{\pi_a^\prime,\pi_b^\prime\}$ pair
and the $\{\pi_a^{\prime\prime},\pi_b^{\prime\prime}\}$ pair
on shell). As we shall illustrate in more detail in the next section, 
this observation allows us to evaluate in a simpler way also the effects 
of $\pi\pi$ scattering in different channels (fig.~\ref{fig:3p}c) and, 
in particular, to express them in terms of the  $\pi\pi$ scattering lengths.

\subsection{Rescattering in \Kp00}
	\begin{figure}[t]
		\begin{center}
		\includegraphics[scale=1]{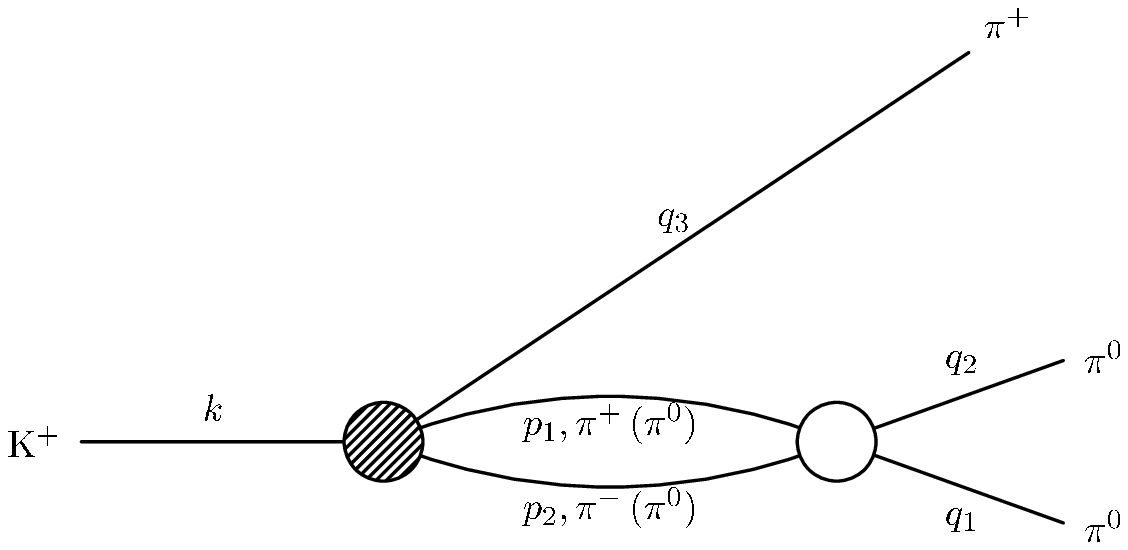}\\ 
		\end{center}
		\caption{The \pppm ( \p00) contribution to \Kp00. \label{fig1}}
	\end{figure}
We start by considering the rescattering of the two-pion pair
leading to the  final \p00 state. In the $\R^{2}$ term of eq. \eqref{R2I} 
the contribution of intermediate \pppm states --- fig.~\ref{fig1} --- is given by
\begin{align}
   \begin{split}
	(\Im \M_{00+})_{\pppm}=\frac{1}{2}\int\frac{d^{3}p_{1}d^{3}p_{2}}{4E_{1}E_{2}}
	&\delta^{4}(p_{1}+p_{2}-q_{1}-q_{2})
	\Re\M_{x}\left((q_{1}+q_{2})^{2}\right) \\
	&\Re \M_{++-}\left((q_{1}+q_{2})^{2},(p_{2}+q_{3})^{2},(p_{1}+q_{3})^{2}\right)\\
   \end{split}
\end{align}
This expression is directly proportional to $v_{\pm}(s_{3})$, so that it will contribute 
to the imaginary part of $B_{00+}$. 
In the next section we will find that the real parts of $D^{1}_{++-}, D^{2}_{++-}$ are 
of the second order in the scattering lengths, so that the contributions of these 
terms to $\Im \M_{00+}$ are $O(a_{i}^{3})$ and can be neglected. 
It is convenient to compute the result in the C.M. of the \p00 pair. 
\begin{figure}[t]
		\begin{center}
		\includegraphics[scale=0.6]{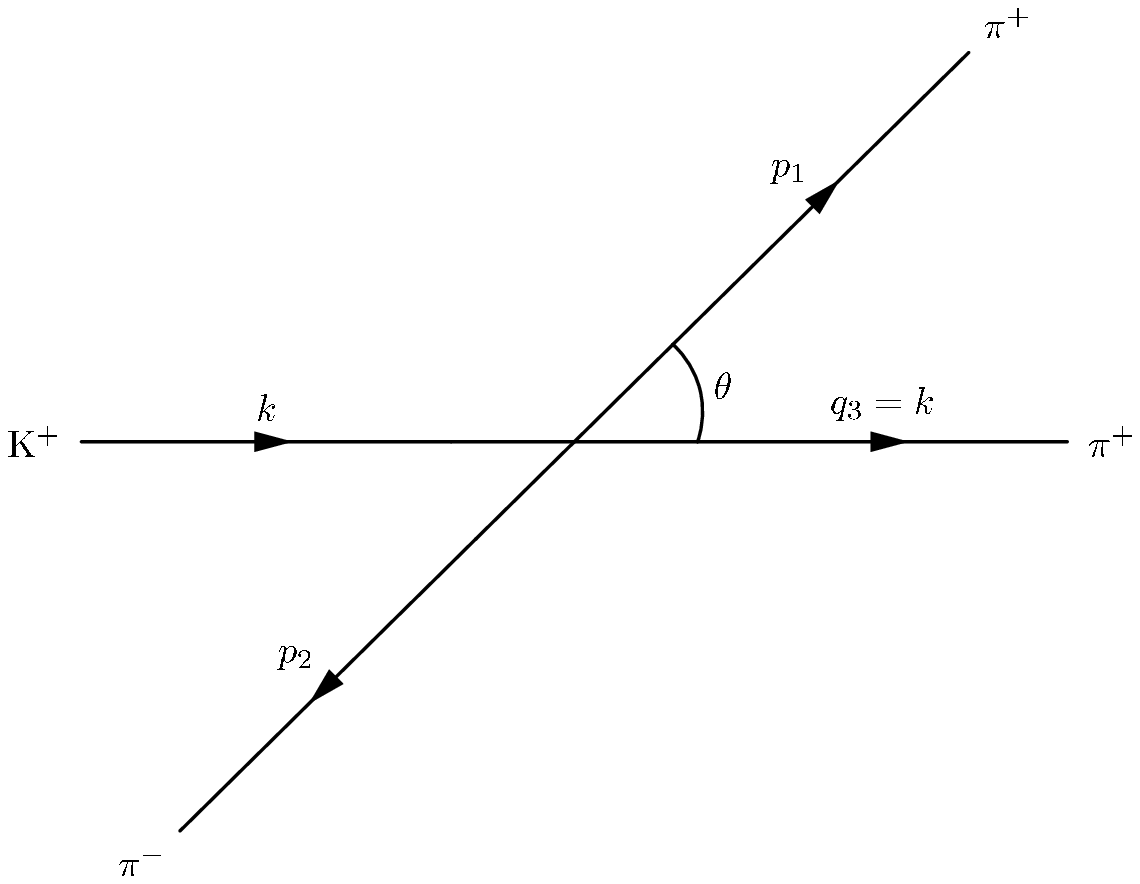}\\ 
		\end{center}
		\caption{The \pppm contribution: kinematics in the \p00 CM. \label{fig2}}
\end{figure}
With reference to fig.~\ref{fig2}, and using eqs. \eqref{Mx},  \eqref{Bx}, we then find 
\begin{align}
	\notag&(\Im B_{00+})_{\pppm}=2 a_{x}(s_{3}) 
		\overline{R^{+}}~, 
\end{align}
 where
\begin{align}
	\overline{R^{+}}&= \frac{1}{2}\int d \cos(\theta)\;
		R^{+}\left((p_{1}+q_{3})^{2}\right) \\
\intertext{To a good accuracy the integrand is linear in $\cos(\theta)$  --- see \eqref{Mplus} ---
so that the average is simply the value at $\theta=\pi/2$. In this approximation we can write}
	\label{imb00}&(\Im B_{00+})_{\pppm}=2 a_{x}(s_{3}) R^{+}(\sbar_3 )
\end{align}
where 
\be
\sbar_i =  \frac{3 s^0_{0}-s_{i}}{2} 
\label{eq:sbar}
\ee
Eq.~(\ref{imb00}) reduces to the result 
in ref. \cite{Cabibbo:2004gq} in the limit $s_{3}\to 4m_{\pi^{+}}^{2}$.
We next consider the contribution of 
intermediate \p00 states (see fig. \ref{fig1}):
\begin{align}
   \begin{split}
	(\Im \M_{00+})_{\p00}=\frac{1}{4}\int\frac{d^{3}p_{1}d^{3}p_{2}}{4E_{1}E_{2}}
	&\delta^{4}(p_{1}+p_{2}-q_{1}-q_{2})
	\Re\M_{00}\left((q_{1}+q_{2})^{2}\right)\\
	&\Re \M_{00+}\left((p_{2}+q_{3})^{2},(p_{1}+q_{3})^{2},(q_{1}+q_{2})^{2}\right)\\
   \end{split}
\end{align}
We can substantially simplify the computation if we neglect the dependence of $\Re\M_{00+}$ on 
$s_{1} $ and $ s_{2}$. We thus obtain
\be
	\label{im00}(\Im \M_{00+})_{\p00}= \frac{\pi v_{00}(s_{3})}{8}\Re\M_{00}(s_{3}) \Re \M_{00+}(s_{3}) 
\ee
We must distinguish the two cases, above and below the \pppm threshold. Above the threshold we obtain
\begin{align}
	(\Im A_{00+})_{\p00}
		&=v_{00}(s_{3})a_{00}(s_{3})R^{0}(s_{3}),  \\
		\label{im00p}(\Im B_{00+})_{\p00}&=v_{00}(s_{3})a_{00}(s_{3})\Re B_{00+}(s_{3})\\
\intertext{Below the \pppm threshold the real parts of $\M_{00+}$ and $\M_{00}$ acquire contributions from the imaginary parts of $B_{00+}$ and of $B_{00}$, see eqs. \eqref{KBelow} and \eqref{PiBelow}, and these 
result in a $O(a_{i}^2)$ contribution to $\Re B_{00+}$,}
	\label{reb00}(\Re B_{00+})_{\p00}
		&=-2v_{00}(s_{3})a_{x}(s_{3})\left[a_{x}(s_{3})R^{0}(s_{3})
			+a_{00}(s_{3})R^{+}(\sbar_{3})\right]
\end{align}
where we have used the results of eqs. \eqref{IB00F} and \eqref{imb00}. 
The presence of a real part of $B_{00+}$ implies an extra contribution to $\Im B_{00+}$, eq. \eqref{im00p}, 
that is however of the third order in the scattering lengths, and can be neglected at $O(\R^{2})$.
 
\begin{figure}[t]

		\begin{center}
		\includegraphics[scale=1]{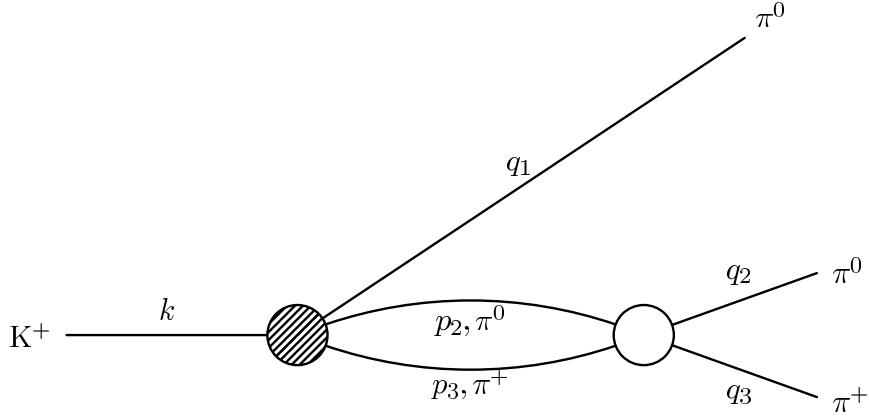}\\ 
		\end{center}
		\caption{The $\pi^{+}\pi^{0}$ contribution. \label{fig3}}
	\end{figure}

Figure~\ref{fig3} shows the two-pion rescattering in one of 
the two $\pi^{+}\pi^{0}$ channels (the other is obtained by the exchange
$q_{1}\leftrightarrow q_{2} $).  Here the situation is 
simpler since there is only one possible intermediate state:
\begin{align}
	\label{pp0ctr}\begin{split}
	(\Im \M_{00+})_{\pi^{0}\pi^{+}}=\frac{1}{2}\int&\frac{d^{3}p_{2}d^{3}p_{3}}{4E_{2}E_{3}}
	\delta^{4}(p_{2}+p_{3}-q_{2}-q_{3})\,
	\Re\M_{+0}\left((q_{2}+q_{3})^{2}\right) \\
	&\quad\quad\quad\quad
	    \Re \M_{00+}\left((q_{2}+q_{3})^{2},(q_{1}+p_{3})^{2},(q_{1}+p_{2})^{2}\right)\\
	&+ (q_{1}\leftrightarrow q_{2})
	\end{split}
\end{align} 
As anticipated, to a good approximation we can neglect the quadratic 
terms in eq. \eqref{Mzero}. In this limit we obtain
\bea
  (\Im A_{00+})_{\pi^{0}\pi^{+}} &=&  (\Im A^{(1)}_{00+})_{\pi^{0}\pi^{+}}  +
  (\Im A^{(2)}_{00+})_{\pi^{0}\pi^{+}}  \nn \\
&\equiv&  (\Im A^{(1)}_{00+})_{\pi^{0}\pi^{+}}~ + ~(s_{1}\leftrightarrow s_{2}) 
\eea
with
\be 
  (\Im A^{(1)}_{00+})_{\pi^{0}\pi^{+}}= 2 a_{+0}(s_{1})v_{+0}(s_{1}) 
 R^{0} \left( \sbar_1 - \Delta_1 \right) 
    +  a_{1} v^3_{+0}(s_1)  \Ac^{0} g^{0} ~ \frac{s_1(s_3-s_2) }{16 m^4_{\pi^+}}~,
\ee
where the ``velocity'' $v_{+0}(s)$ and  $\Delta_i$ are defined by 
\bea
v_{+0}(s) &=& \frac{\left( s-(m_{\pi^{+}}+m_{\pi^{0}})^{2}\right)^{1/2}
			\left( s-(m_{\pi^{+}}-m_{\pi^{0}})^{2}\right)^{1/2}}{s}  \\
\Delta_i  &=&  \frac{(m_{\pi^+}^2 -m_{\pi^0}^2)( m_{K}^2 -m_{\pi^0}^2 )}{4 s_i } 
\eea

\medskip

Finally, we must take into account the effective $3\pi\to3\pi$ 
scattering due to reducible diagrams of the type in fig.~\ref{fig:3p}c. 
By construction, these contributions are at least of $O(a_i^2)$
and at this level of accuracy contribute only to the 
real part of the amplitude. Following the decomposition 
in eq.~(\ref{KAbove})--(\ref{KBelow}), the various rescattering 
combinations of the type in fig.~\ref{fig:3p}c can be 
divided into two main groups: those which can be reabsorbed 
into a redefinition of $\Re A_{00+}$ and those which 
affect $\Re B_{00+}$. We shall start from the latter,
that are more relevant for the structure of the cusp.

The $O(a_i^2)$ corrections to  $\Re B_{00+}$ arise from 
diagrams  of the type in fig.~\ref{fig:3p}c
with the identification $\{\pi_a,\pi_b,\pi_c\} \equiv \{\pi^+,\pi^0,\pi^0 \}$ and 
$\{\pi^{\prime\prime}_b,\pi^\prime_c\} \equiv \{\pi^\pm\pi^\mp\}$.
We can express all these contributions as
\bea
(\Re B_{00+})_{{\rm fig}.\,\ref{fig:3p}c}
&=&  -2 a_{x}(s_{3}) \times \left[  (\Im C_{++-})_{\pppp}(\sbar_3)  \right. \nn \\
&& \left. + (\Im C^{(1)}_{++-})_{\pppm}(\sbar_3) 
   + v_{00}(\sbar_3) (\Im D^{(1)}_{++-})_{\p00} (\sbar_3) \right] \label{eq:2r1}
\eea
As we shall see in the next section, the three terms in the r.h.s. of eq.~\eqref{eq:2r1}
correspond to the cases 
where the $\{\pi^\prime_a,\pi^\prime_b\}$ pair is identified with 
$\{\pi^+,\pi^+\}$, $\{\pi^+,\pi^-\}$, or $\{\pi^0,\pi^0\}$. 
Summing these three terms we find 
\bea
(\Re B_{00+})_{{\rm fig}.\,\ref{fig:3p}c}
&=&  -2a_{x}(s_{3}) \left[  a_{++}(\sbar_3 )  v_\pm(\sbar_3)  R^+(\sbar_3 ) \right.  \nn \\
&& \left. 
  + 2 a_{+-}( \sbar_3 ) v_{\pm}(\sbar_3 ) R^{+}( \langle\sbar\rangle_3 ) 
  + a_{x}(\sbar_3) v_{00}(\sbar_3 ) R^{0}(\sbar_3)  \right] \nn \\
\eea
where $\langle\sbar\rangle_3 = (3s_0+s_3)/4$ and, given we are 
already at $O(a_i^2)$, we have neglected the tiny  P-wave contribution
and the difference between $s_0^0$ and $s_0^+$ in the $\sbar_i$
variables.

{}Far from the Dalitz plot boundaries, the $O(a_i^2)$ corrections to  
$\Re A_{00+}$ could be ignored since they can be reabsorbed into a
redefinition of $R^0(s)$. However, the polynomial form of $R^0(s)$
is not appropriate to describe the square-root singularities that occur 
at the borders of the Dalitz plot and, particularly, at 
$\pi^0\pi^0$ and $\pi^+\pi^0$ thresholds. The latter are
described at $O(a_i^2)$ accuracy by the remaining diagrams 
of the type in fig.~\ref{fig:3p}c. 
The singularities at the $\pi^0\pi^0$ threshold, that  are obtained with 
the identification  $\{\pi^{\prime\prime}_b,\pi^\prime_c\} \equiv \{\pi^0\pi^0\}$,
are
\bea
 (\delta \Re A_{00+})_{{\rm fig}.~\ref{fig:3p}c}^{\pi^0\pi^0}
&=&  -a_{00}(s_{3}) v_{00} (s_3) 
\times \left[  (\Im A^{(1)}_{+00})_{\pp0}(\sbar_3)  + (\Im A^{(2)}_{+00})_{\pp0}(\sbar_3) 
\right] \nn \\
&& = -4 a_{00}(s_{3})  v_{00} (s_3) a_{+0}(\sbar_3 )v_{+0}( \sbar_3 ) 
 R^{0} (\langle\sbar\rangle_3) 
\eea
where again we have neglected the tiny  P-wave contribution
and the difference between $\pi^0$ and $\pi^+$ masses 
in the $\sbar_i$ variables.
The singularities at the $\pi^+\pi^0$ thresholds,
obtained with  $\{\pi^{\prime\prime}_b,\pi^\prime_c\} \equiv \{\pi^+\pi^0\}$
or $\{\pi^0\pi^+\}$, are 
\bea
(\delta \Re A_{00+})_{{\rm fig}.~\ref{fig:3p}c}^{\pi^+\pi^0}
&=&  -2a_{+0}(s_{1}) v_{+0} (s_1) 
\times \left[  (\Im A^{(2)}_{+00})_{\pp0}(\sbar_1) + (\Im A_{+00})_{\p00}(\sbar_1)\right. \nn \\
&& \left.  + v_\pm(\sbar_1) (\Im B_{+00})_{\pppm}(\sbar_1) \right] ~ + ~(s_{1}\leftrightarrow s_{2}) 
\nn \\
&=& -2a_{+0}(s_{1}) v_{+0} (s_1)  
\left[ 2a_{+0}(\sbar_1) v_{+0}(\sbar_1) R^0(\langle\sbar\rangle_1)  \right. \nn \\
 && \left. + a_{00}(\sbar_1) v_{00}(\sbar_1) R^0(\sbar_1)
 + 2 a_{x}(\sbar_1) v_{+-}(\sbar_1) R^+(\sbar_1)  \right] 
\nn \\  &&  ~ + ~(s_{1}\leftrightarrow s_{2})
\eea

\medskip    

In summary, the relevant contribution to the 
$\Kp00$ amplitude (in addition to $\Re A_{00+}$) are:
\bea
\Im B_{00+} &=&2 a_{x}(s_{3}) R^{+}(\sbar_3), \label{eq:Kp00_1} \\
\Re B_{00+} &=& -2a_{x}(s_{3}) \left[ 
  a_{x}(s_{3})  v_{00}(s_3) R^0(s_3)  + a_{00}(s_{3})  v_{00}(s_3) R^+(\sbar_3)
\right. \nn \\ &&  \left.
+ a_{++}(\sbar_3 )  v_\pm(\sbar_3)  R^+(\sbar_3 ) 
+ 2 a_{+-}( \sbar_3 )
	v_{\pm}(\sbar_3 ) R^{+}(\langle\sbar\rangle_3) \right. \nn \\ &&  \left.
+  a_{x}(\sbar_3) v_{00}(\sbar_3 ) R^{0}(\sbar_3)  
\right]  \\
\Im A_{00+} &=& 
  a_{00}(s_{3}) v_{00}(s_{3}) R^0(s_3) + \Big[ 
  2 a_{+0}(s_{1})v_{+0}(s_{1}) R^{0} \left( \sbar_1 - \Delta_1 \right)  \nn \\
    && \qquad   +   a_{1} v^3_{+0}(s_1)  \Ac^{0} g^{0} ~ \frac{s_1(s_3-s_2) }{16 m^4_{\pi^+}} ~
+ ~ (s_{1}\leftrightarrow s_{2}) \Big] \label{eq:Kp00_2} 
\eea
In order to take into account also the $O(a_i^2)$ 
singularities at the Dalitz-plot boundaries, 
$\Re A_{00+}$ must be modified with 
the addition of the following extra term:
\be
 \Re A_{00+} \to R^0(s_3) + \delta \Re A_{00+}
\ee
\bea
 \delta \Re A_{00+} &=& 
 -4 a_{00}(s_{3})  v_{00} (s_3) a_{+0}(\sbar_3 )v_{+0}( \sbar_3 ) \nn \\
&& -\Big\{2a_{+0}(s_{1}) v_{+0} (s_1) \big[ 2a_{+0}(\sbar_1) v_{+0}(\sbar_1) R^0(\langle\sbar\rangle_1) 
    \nn \\
 && \qquad + a_{00}(\sbar_1) v_{00}(\sbar_1) R^0(\sbar_1)  \nn \\
 && \qquad + 2 a_{x}(\sbar_1) v_{+-}(\sbar_1) R^+(\sbar_1)  \big] 
~ + ~(s_{1}\leftrightarrow s_{2}) \Big\} \label{DPbound}
\eea

\subsection{Rescattering in \Kppm}
In evaluating the coefficient of the cusp for the \Kp00 decay we need 
to extract from data the real part of the \Kppm amplitude. 
We thus need a suitable parameterization of the latter 
at the same level of accuracy. Since in the  \Kppm case the
physical region is always above threshold, we do not expect 
any correction at $O(a_i)$ in the decay distribution. 
This implies that the parametrization \eqref{Mplus} 
for $\Re C_{++-}$ determines the  
real part of the \Kppm decay amplitude at $O(a_i)$ accuracy. 
Since the \Kppm amplitude appears only multiplied by $O(a_i)$   
coefficients in the \Kp00 rate, knowing the \Kppm amplitude
at  $O(a_i)$ accuracy is sufficient to the purpose of evaluating the cusp 
effect in \Kp00 at the $O(a^2_i)$ level.

However, it is worth to stress that the $O(a^2_i)$ corrections 
to the \Kppm decay amplitude have their own interest:
at the border of the Dalitz plot they give rise to square-root 
singularities that could eventually be detected.
Their inclusion would therefore improve the quality 
of the \Kppm parameterization and could even be used to extract 
an additional information about the \pipi re-scattering.
For this reason,  in the following we 
shall provide a complete parameterization of the 
re-scattering effects in the  \Kppm amplitude to $O(a^2_i)$ accuracy.

We start from the expression in eq.~\eqref{Kppm}
and the parametrization \eqref{Mplus}, where the 
notation of the momenta is defined by 
\begin{equation*}
	K^{+}\to \pi^{+}(q_{1})+\pi^{+}(q_{2})+\pi^{-}(q_{3});
	\quad\quad s_{1}=(q_{2}+q_{3})^{2}, \text{  etc.}
\end{equation*}
In analogy with the \Kp00  case, we decompose the amplitude
isolating explicitly the cusp effect related to the 
$\pi^0\pi^0 \leftrightarrow \pi^+\pi^-$ transition. 
In the physical case, where $m_{\pi^{+}} > m_{\pi^{0}}$, this cusp effect 
is not observable; however, the corresponding amplitude 
is still well defined. As discussed in section \ref{sect:masscont},
this cusp amplitude is more conveniently analysed in the  unphysical 
scenario with $m_{\pi^{0}}>m_{\pi^{+}}$.
In this scenario the \p00 threshold gives rise to two square-root singularities, 
respectively in $s_{1}$ and in $s_{2}$. 
To cover the case where either $s_{1}$ or $s_{2}$ are below the respective threshold, 
eq.~\eqref{Kppm} must be completed as follows:  
\begin{align}
    \M_{++-} &= C_{++-} + D^{(1)}_{++-}v_{00}(s_{1})+ D^{(2)}_{++-}v_{00}(s_{2})
    	&& s_{1,2}>4m_{\pi^{0}}^{2},
    \label{KPMAbove}\\
  \M_{++-} &= C_{++-} + iD^{(1)}_{++-}v_{00}(s_{1})+ D^{(2)}_{++-}v_{00}(s_{2})
  	&& s_{1}<4m_{\pi^{0}}^{2}
    \label{KPMBelows1},\\
  \M_{++-} &= C_{++-} + D^{(1)}_{++-}v_{00}(s_{1})+ iD^{(2)}_{++-}v_{00}(s_{2})
  	&& s_{2}<4m_{\pi^{0}}^{2}
   \label{KPMBelows2}
\end{align}  
We can choose $m_{\pi^{0}}$ close to $m_{\pi^{+}}$ so that $s_{1}$ and  $s_{2}$ 
cannot simultaneously be below the \p00 threshold.

As far as the two-pion scattering is concerned, 
we must take into account \p00, \pppm and \pppp intermediate states. 
In computing the \p00 contribution to $D^{(1)}_{++-}$ and $D^{(2)}_{++-}$  
we will assume that $\Re A_{00+}$ is only a function of $s_{3}$ as in the parametrization  
\eqref{Mzero}, and obtain 
\be
	\label{imDp00}(\Im D^{(1,2)}_{++-})_{\p00}= a_{x}(s_{1,2}) 
		R^{0}(s_{1,2})
\ee
Analogously, for the \pppp intermediate state we obtain
\be
	\label{imCppp}(\Im C_{++-})_{\pppp}= a_{++}(s_{3}) v_{\pm}(s_{3})
		R^{+}(s_{3})
\ee
We next consider the contribution of \pppm\ra\pppm rescattering, whose general expression is.  
\begin{align}
	\label{ppmctr}\begin{split}
	(\Im \M_{++-})_{\pppm}=\frac{1}{2}\int&\frac{d^{3}p_{2}d^{3}p_{3}}{4E_{2}E_{3}}
	\delta^{4}(p_{2}+p_{3}-q_{2}-q_{3})\,
	\Re\M_{+-}\left((q_{2}+q_{3})^{2}\right) \\
	&\quad\quad\quad\quad
	    \Re \M_{++-}\left((q_{2}+q_{3})^{2}, (q_{1}+p_{3})^{2}, 	
	    			(q_{1}+p_{2})^{2}\right)\\
	&+ (q_{1}\leftrightarrow q_{2})
	\end{split}
\end{align} 
Above either of the \p00 thresholds\footnote{~We recall  that we are considering the situation where $\mpo>\mpp$.} 
in $s_{1}$ and $s_{2}$, we can neglect the contributions of the real parts of $D^{1,2}_{++-}$ to $\Re \M_{++-}$ that, as we will shortly see, are $O(a_{i}^{2})$ and give a $O(a_{i}^{3})$ contribution to \eqref{ppmctr}. Using the parametrization \eqref{Mplus} and considering only S-wave scattering leads to 
\be
(\Im C_{++-})_{\pppm} = (\Im C^{(1)}_{++-})_{\pppm} + (\Im C^{(2)}_{++-})_{\pppm}
\ee 
with
\be
 (\Im C^{(1,2)}_{++-})_{\pppm} = 2 a_{+-}(s_{1,2})
	v_{\pm}(s_{1,2}) R^+(\sbar_{1,2}) 
\label{imCppm}
\ee
Note that, although for simplicity of notations we use the same symbol 
adopted in eq.~(\ref{eq:sbar}), in the \Kppm case the $\sbar_i$ variables 
are defined by 
\be
\sbar_i =  \frac{3 s^0_{+}-s_{i}}{2} 
\label{eq:sbar2}
\ee
i.e.~we must replace $s^0_{0}$ with $s^0_{+}$ with respect to
eq.~(\ref{eq:sbar}). If we take into account also the tiny P-wave contribution, 
the above result is modified as follows
\bea
	\label{imCppm_P}(\Im C^{(1)}_{++-})_{\pppm} &=& 2 a_{+-}(s_{1})
	v_{\pm}(s_{1}) R^+(\sbar_1)\,+  a_{1} v^3_{\pm}(s_1) 
    \Ac^{+} g^{+} ~ \frac{s_1(s_3-s_2) }{16 m^4_{\pi^+}} 
\eea

At this point we could consider the (unphysical) case where $s_{1}<4m_{\pi^{0}}^{2}$. 
Here using eq.~\eqref{KPMBelows1} we see that the real part of
$\M_{++-}$ also receives a contribution from the \emph{imaginary} part of $D^{(1)}_{++-}$ that, when injected in  \eqref{ppmctr} produces a contribution to the \emph{real} part of $D^{(1)}_{++-}$. Another contribution to  $\Re D^{(1)}_{++-}$ below the \p00 threshold arises from the $\Im D_{\pm}$ term in $\Re\M_{+-}$, see eqs. \eqref{PMBelow} and \eqref{IDpmF}. 
Both these contributions are  $O(a_{i}^{2})$ and lead to
\begin{align}
	\label{ppmrectr}\begin{split}
	(\Re D^{(1)}_{++-})_{\pppm}
		&=-\frac{4}{\pi}\int\frac{d^{3}p_{2}d^{3}p_{3}}{4E_{2}E_{3}}
		\delta^{4}(p_{2}+p_{3}-q_{2}-q_{3})\,\\
		&\quad\quad\left(R^{+}((q_{1}+p_{2})^{2}) a_{x}^{2}(s_{1})+R^{0}(s_{1}) 
			a_{x}(s_{1})a_{\pm}(s_{1})\right)
	\end{split}\\
		&=-2 v_{\pm}(s_{1})\left[ R^{+}(\sbar_1) a_{x}^{2}(s_{1})+R^{0}(s_{1}) 
			a_{x}(s_{1})a_{\pm}(s_{1}) \right]
\end{align} 

Finally, we must take into account the effective $3\pi\to3\pi$ 
scattering due to reducible diagrams of the type in fig.~\ref{fig:3p}c. 
As in the $K^+\to\pi^+\pi^0\pi^0$ case, these contributions are of $O(a_i^2)$
and contribute only to the real part of the amplitude. 
The $O(a_i^2)$ corrections to  $D^{(1,2)}_{++-}$ are 
\bea
(\Re D^{(1)}_{++-})_{{\rm fig}.~\ref{fig:3p}c} &=& - a_{x}(s_{1}) 
\times \left[  (\Im A^{(1)}_{00+})_{\pi^{0}\pi^{+}}(\sbar_1) + 
 (\Im A^{(2)}_{00+})_{\pi^{0}\pi^{+}}(\sbar_1) \right] 
\nn \\
   &=&  - 4 a_{x}(s_{1}) a_{+0}(\sbar_{1})v_{+0}(\sbar_{1}) 
 R^{0} \left( \langle\sbar\rangle_1 \right) 
\eea
While the remaining  $O(a_i^2)$ corrections, that can be absorbed 
into a redefinition of the $\Re C_{++-}$, are 
\bea
 (\delta \Re C_{++-})_{{\rm fig}.~\ref{fig:3p}c}^{\pi^+\pi^+}
&=&  -a_{++}(s_{3}) v_{\pm} (s_3) 
\times \left[  (\Im C^{(1)}_{++-})_{\pppm}(\sbar_3)  
 +  (\Im C^{(2)}_{++-})_{\pppm}(\sbar_3) \right. \nn \\
&& \left. + v_{00}(\sbar_3) (\Im D^{(1)}_{++-})_{\p00}(\sbar_3)  
 + v_{00}(\sbar_3) (\Im D^{(2)}_{++-})_{\p00}(\sbar_3) \right] \nn \\
&=& - a_{++}(s_{3})  v_{\pm} (s_3) \left[ 2  a_x(\sbar_3) v_{00}(\sbar_3) R^0(\sbar_3) \right. \nn \\
&& \left. + 4 a_{+-}(\sbar_3) v_\pm(\sbar_3) R^{+} (\langle\sbar\rangle_3) \right]
\eea
and 
\bea
(\delta \Re C_{++-})_{{\rm fig}.~\ref{fig:3p}c}^{\pi^+\pi^-}
&=&  -2a_{+-}(s_{1}) v_{\pm} (s_1) 
\times \left[  (\Im C_{++-})_{\pi^+\pi^+}(\sbar_1) + (\Im C^{(2)}_{++-})_{\pppm}(\sbar_1)\right. \nn \\
&& \left.  + v_{00}(\sbar_1) (\Im D^{(2)}_{++-})_{\pppm}(\sbar_1) \right] ~ + ~(s_{1}\leftrightarrow s_{2}) 
\nn \\
&=& -2a_{+-}(s_{1}) v_{+-} (s_1)  
\left[ 2a_{+-}(\sbar_1) v_{+-}(\sbar_1) R^+(\sbar_1)  \right. \nn \\
 && \left. + 2 a_{+-}(\sbar_1) v_{+-}(\sbar_1) R^+(\langle\sbar\rangle_1)
 + a_{x}(\sbar_1) v_{00}(\sbar_1) R^0(\sbar_1)  \right] 
\nn \\  &&  ~ + ~(s_{1}\leftrightarrow s_{2})
\eea

In summary, the relevant contributions to the 
\Kppm amplitude (in addition to $\Re C_{++-}$) are:
\bea
\Im D^{(1)}_{++-} &=& a_{x}(s_{1}) R^{0}(s_{1}), \label{eq:Kppm_1} \\
\Re D^{(1)}_{++-} &=& -2 a_{x}(s_{1}) \left[  v_{\pm}(s_{1}) R^{+}(\sbar_1) a_{x}(s_{1})
 + v_{\pm}(s_{1}) R^{0}(s_{1}) a_{\pm}(s_{1})  \right. \nn \\ && 
 \left.       + 2 v_{+0}(\sbar_{1}) a_{+0}(\sbar_{1})  R^{0} 
\left( \langle\sbar\rangle_1 \right) \right]  \\
\Im C_{++-} &=& a_{++}(s_{3}) v_{\pm}(s_{3})
		R^{+}(s_{3}) +\Big[ 2 a_{+-}(s_{1}) v_{\pm}(s_{1}) R^+(\sbar_1)\nn \\ && 
 +  a_{1} v^3_{\pm}(s_1) \Ac^{+} g^{+} ~ \frac{s_1(s_3-s_2) }{16 m^4_{\pi^+}} 
 + ~ (s_{1}\leftrightarrow s_{2}) \Big] \label{eq:Kppm_2} 
\eea

\subsection{The $K_L \to 3\pi$ system}

The two $K_L \to 3\pi$ coupled channels
form a system very similar to the one of the two $K^+ \to 3\pi$ decays.
Similarly to the charged modes, we can decompose the two $K_L$
decay amplitudes into regular terms and terms that are  
singular at the \pppm (\p00) threshold: 
\bea
   \M_{000} &=& A_{000} + \sum_{i=1\ldots3} B^{(i)}_{000}  v_\pm(s_i)
   \left[ \Theta(s_i - 4m_{\pi^{+}}^{2}) +i \Theta(4m_{\pi^{+}}^{2}-s_i)\right]  ~, 
    \label{K000}\\
   \M_{+-0} &=& C_{+-0} + D_{++0} v_{00}(s_3)
   \left[ \Theta(s_3 - 4m_{\pi^{0}}^{2}) +i \Theta(4m_{\pi^{3}}^{2}-s_i)\right]  ~.
\eea
Concerning the leading amplitudes ($\Re A_{000}$ and 
$\Re C_{+-0}$) we shall adopt the following phenomenological 
parametrization:
\begin{align}
 	\Re A_{000}(s_{1},s_{2},s_{3}) &= R_L^{0}(s_1,s_2,s_{3}) 
	= \AL^{0} \left[ 1+ \tilde h_L^{0} \sum_{i=1\ldots3} (s_{i}-s^0_{0_L})^2/3\mpp^{4}
	 	  \right] \label{MzeroL}\\
 	\Re C_{+-0}(s_{1},s_{2},s_{3}) &= R_L^{+}\!(s_{3}) 
	= \AL^{+} \left[ 1+g_L^{+}\!(s_{3}-s^+_{0_L})/2\mpp^{2}
		+\tilde h_L^{+}\!(s_{3}-s^+_{0_L})^{2}/2\mpp^{4} \right] \label{MplusL}	
\end{align}
where 
\be 
\sum_{i=1\ldots3} s_i = \left\{ \ba{ll} 
3 s_{0_L}^0  = m_K^2 + 3 m^2_{\pi^0}                    &(K_L \to 3 \pi^0 ) \\
3 s_{0_L}^+    = m_K^2 + 2 m^2_{\pi^+} + m_{\pi^0}  \quad &( K_L \to \pi^+\pi^-\pi^0) \ea \right.
\ee
The values of the  $K_L \to \pi^+\pi^-\pi^0$ slopes fitted by PDG
(that also includes a small term proportional to $(s_1-s_2)^2$)
are $g_L^{+} \approx 0.678\pm 0.008$ and 
$\tilde h_L^{+} \approx h_{\text{\tiny PDG}} -( g_L^+/2 )^{2} = 0.04 \pm 0.01$.
In the $K_L \to 3\pi^0$ case the linear term is 
forbidden by Bose symmetry; the normalization of the quadratic term 
has been chosen such that $\tilde h_L^{0} \approx h_{\text{\tiny PDG}}
\approx - (5.0 \pm 1.4)\times 10^{-3}$.

Here the visible cusp due to the $\pi^+\pi^- \to \pi^0\pi^0$ 
rescattering is expected in the $K_L \to 3\pi^0$ spectrum.
The phenomenon is completely analog to what happens in
the charged modes; however, the relative size of the 
cusp is smaller because of the inverted hierarchy 
in the leading amplitudes: in the isospin limit 
$\AL^+/\AL^{0}=1/3$, to be compared with the 
analogous ratio $\Ac^+/\Ac^{0}=2$ of the charged modes.

The calculation of the imaginary parts of the amplitudes 
(and the real part of the cusp coefficient) proceeds 
exactly as in the charged modes. We report here only 
the results. In the interesting case of the 
$K_L \to 3\pi^0$ amplitude, 
the imaginary parts are 
\bea
(\Im B^{(i)}_{000})_{\pppm} &=& 2 a_{x}(s_{i}) R_L^{+}(s_{i})~,  \\
(\Im A_{000})_{\p00} &=& \sum_{i=1\ldots3} a_{00}(s_{i})  v_{00}(s_i) 
 R_L^{0}(s_{i},\sbar_i,\sbar_i)~, 
\eea
with
\be
\sbar_i =  \frac{3 s^0_{0_L}-s_{i}}{2}~,
\label{eq:sbar_2}
\ee
while the $O(a_i^2)$ corrections to the (visible) cusp amplitude are
\bea
(\Re B^{(i)}_{000})_{{\rm fig}.~\ref{fig:3p}a} &=&
-2v_{00}(s_{i})a_{x}(s_{i})\left[a_{x}(s_{i}) R_L^0(s_{i},\sbar_i,\sbar_i)
			+a_{00}(s_{i})R^{+}(s_{i})\right]~, \\
(\Re B^{(i)}_{000})_{{\rm fig}.~\ref{fig:3p}c}
&=&  -8 a_{x}(s_{i}) a_{+0}(\sbar_i) v_{+0}(\sbar_i) R^{+}(\langle\sbar\rangle_i)~.
\eea
For the auxiliary mode,  $K_L \to \pi^+\pi^-\pi^0$,  we find 
\bea
(\Im D_{+-0})_{\p00} &=&  a_{x}(s_{3}) R_L^{0}(s_{3},\sbar_3,\sbar_3)~,  \\
(\Im C_{+-0})_{\pi^\pm \pi^0}  &=& 2 a_{+0}(s_{1})v_{+0}(s_{1}) 
 R_L^{+} \left( \sbar_1 + \Delta_{1_L} \right) \nn \\
    & + &  a_{1} v^3_{+0}(s_1)  \AL^{+} g_L^{+} ~ \frac{s_1(s_3-s_2) }{16 m^4_{\pi^+}} ~
+ ~ (s_{1}\leftrightarrow s_{2})~,
\eea
with 
\be
\Delta_{i_L} =  \frac{(m_{\pi^+}^2 -m_{\pi^0}^2)( m_{K}^2 -m_{\pi^+}^2 )}{4 s_i } 
\ee

\subsection{Decay rates and extraction of the scattering lengths}
The aim of this paper is eminently practical. Our goal is  
\begin{itemize}
\item to establish a parametrization of $K\to 3 \pi$ amplitudes 
      suitable to fit the experimental decay distributions at the 10$^{-3}$ level; 
\item describe the cusp effect due to the $\pppm\to\p00$ rescattering 
      with a theoretical error of a few \%.
\end{itemize}
Since the cusp effect on the rate is $\sim 10\%$,  
the two requests are compatible. In this section we shall outline 
the basic strategy for the extraction of
the combination of scattering lengths $a_x$, as defined in section~\ref{sub:scatt_def},
from a fit to the \Kp00 decay distribution.

\begin{figure}[p]
	\begin{center}
        \vspace{-2.0 cm}
        \includegraphics[scale=0.75]{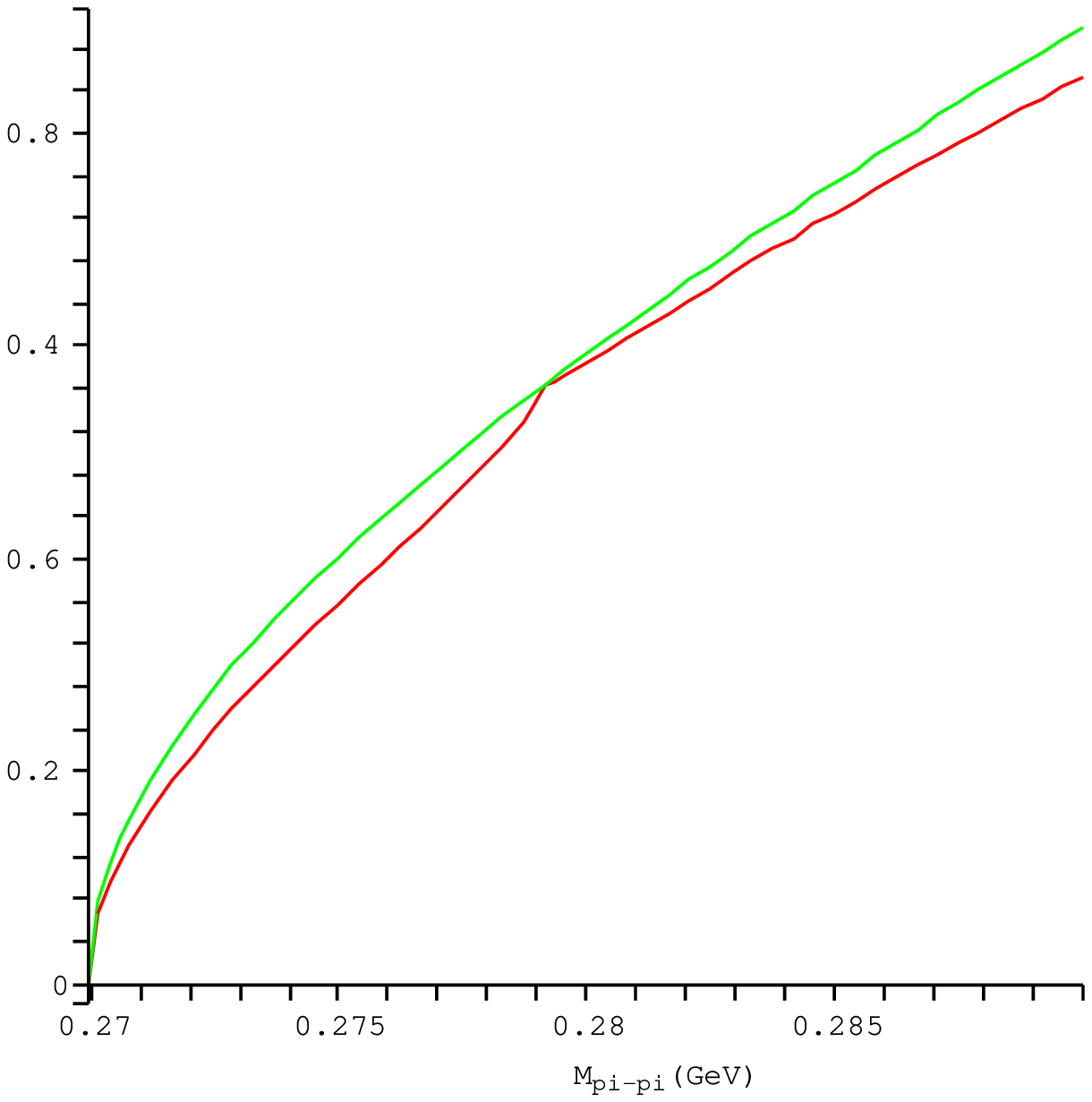} \\
	\vspace{0.0 cm}
        \includegraphics[scale=0.75]{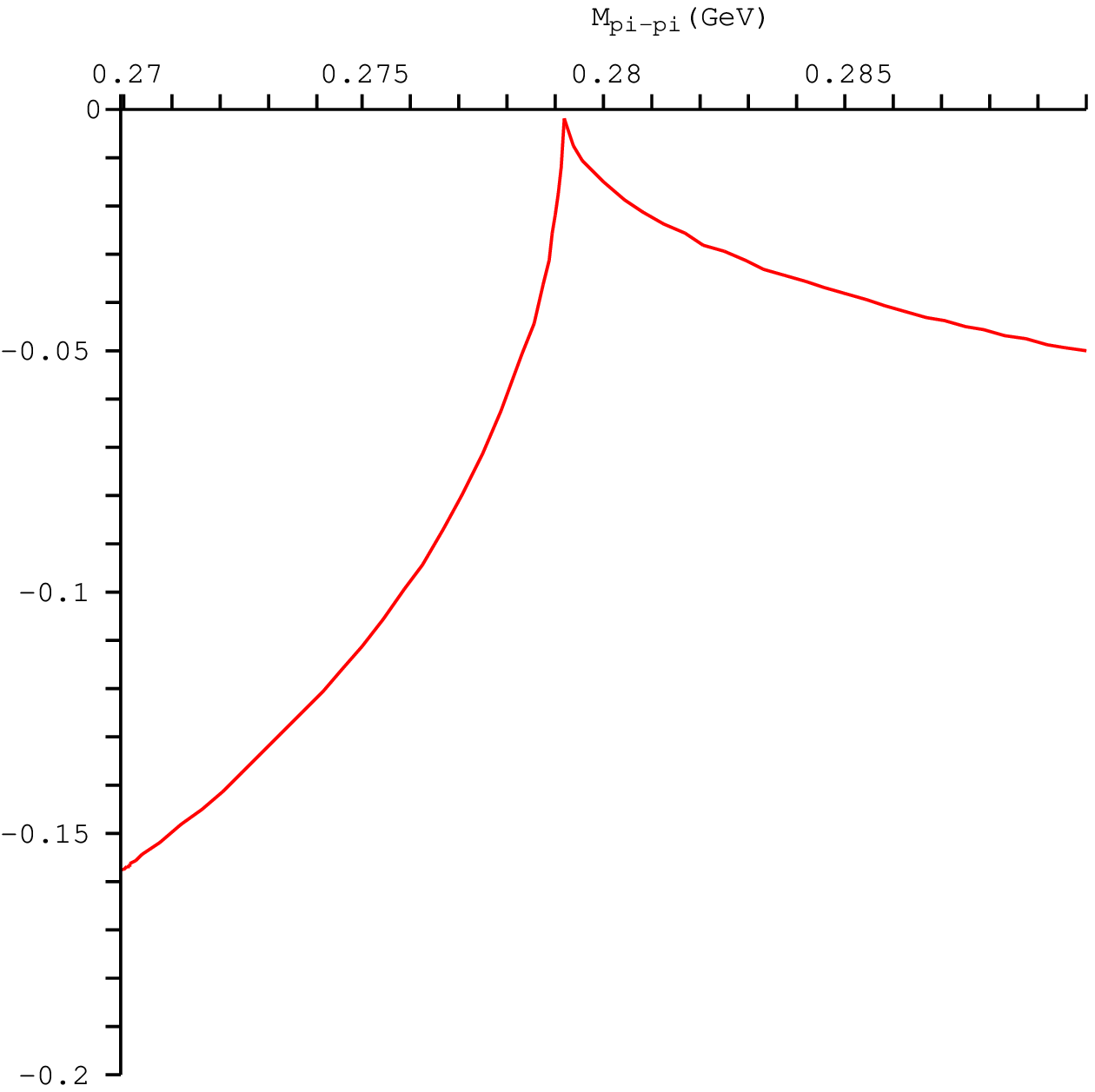} 
	\vspace{-1.0 cm}
        \end{center}   
	\caption{Illustration of the cusp effects in \Kp00. Upper plot: differential decay
distribution (in arbitrary units, as a function of the $\pi^0\pi^0$ invariant mass)
with and without the cusp amplitude.
Lower plot: relative size of the cusp amplitude with respect to the  
regular term. All plots have been obtained using the values of the 
scattering lengths and their effective ranges from Ref.~\cite{Colangelo}
(see sect.~\ref{sub:scatt_def}). \label{fig:plot2} }
\end{figure}

{}From eqs.~\eqref{KAbove}--\eqref{KBelow} the differential decay rate for \Kp00 is
\begin{align}
   | \M_{00+}|^{2} &= \left(\Re A_{00+} + \Re B_{00+}v_{\pm}(s_3)\right)^{2}
   				+  \left(\Im A_{00+} + \Im B_{00+}v_{\pm}(s_3)\right)^{2}
&& s_{3}>4m_{\pi^{+}}^{2} \nn \\
   | \M_{00+}|^{2} &= \left(\Re A_{00+} - \Im B_{00+}v_{\pm}(s_3)\right)^{2}
   				+  \left(\Im A_{00+} + \Re B_{00+}v_{\pm}(s_3)\right)^{2}
&& s_{3}<4m_{\pi^{+}}^{2} \nn 
\end{align}
Expanding the various terms up to $O(a_i^2)$, we can write 
\be
 | \M_{00+}|^{2} = (\Re A_{00+})^2 + \Delta_{A} + v_{\pm}(s_3) \Delta_{\rm cusp} 
 + O(a_i^3)
\ee
where 
\bea
 \Delta_{A} &=& (\Im A_{00+})^2 + v^2_{\pm}(s_3) (\Im B_{00+})^2  \\
\Delta_{\rm cusp} &=&  \left\{ \ba{ll}  
      -2 \Re A_{00+}\Im B_{00+} & s_{3}<4m_{\pi^{+}}^{2} \\
      2 \Re A_{00+}\Re B_{00+} +  2 \Im A_{00+}\Im B_{00+} \quad 
& s_{3} > 4m_{\pi^{+}}^{2} \ea \right.
\label{eq:Dcusp}
\eea
The explicit expressions for the various 
terms are given in eqs.~(\ref{eq:Kp00_1})--(\ref{eq:Kp00_2}).
At the same level of accuracy, the \Kppm decay distribution 
in the physical region is 
\be
 | \M_{++-}|^{2} = (\Re C_{++-})^2 + \Delta_{C} + O(a_i^3) 
\ee
with the $O(a_i^2)$ correction given by 
\bea
\Delta_{C} &=&  \left[ \Im C_{++-} + v_{00}(s_1)\Im D^{(1)}_{++-} 
                        + v_{00}(s_2)\Im D^{(2)}_{++-} \right]^2  \nn \\
&+ & 2 \Re C_{++-} \left[  v_{00}(s_1)\Re D^{(1)}_{++-} 
                        + v_{00}(s_2)\Re D^{(2)}_{++-} \right]
\eea
and the explicit expressions for the various terms 
reported in eqs.~(\ref{eq:Kppm_1})--(\ref{eq:Kppm_2}).

The cusp amplitude in eq.~(\ref{eq:Dcusp}) contains both
a leading $O(a_i)$ term responsible for the negative  
square-root behavior of the rate 
below the threshold and an $O(a^2_i)$ term 
that leads to a similar (smaller) behavior also 
above the threshold (see figure~\ref{fig:plot2}). 
Both these effects are proportional to $a_x$.

The precision with which the coupling $a_x$ can be extracted from data 
depends on the accuracy of our parametrization of the amplitudes and,
in particular, on the theoretical expression for $\Delta_{\rm cusp}$.
Since we have neglected  $O(a_i^3)$ terms, a priori we should expect relative 
corrections of $O(a_i^2)$ to the value of $a_x$. Given the 
expected value of the scattering lengths and the effect of the 
$O(a^2_i)$ terms in figure~\ref{fig:plot2}, a natural 
estimate of this error is about $5\%$. A posteriori checks 
about the size of this error can be obtained 
by studying the stability of the central value of  $a_x$   
obtained by means of different fitting procedures. 
In particular, it would be interesting to compare 
the results obtained under the  following assumptions: 
\begin{enumerate}
\item{} All the $a_i$ are treated as free parameters.
\item{} All the $a_i$ but $a_x$ are fixed to their standard  
        values and only $a_x$ is treated as a free parameter.
\item{} The fit is extended up to the border of the Dalitz Plot with the 
        inclusion of the $\delta \Re A_{00+}$ term in (\ref{DPbound}).
\item{} The expressions of $R^{+,0}(s)$ are modified with the 
        inclusion of cubic  terms in $(s_3-s_0)$ and/or 
        quadratic terms in $(s_1-s_2)$. 
\item{} One of the two $\Delta_{A,C}$ terms (or both) is ignored 
        [in this way the 
        regular amplitudes are re-defined by corrections of $O(a_i^2)$; 
        this, in turn, implies an $O(a_i^2)$ effect on the extraction of 
	$a_x$, of the same order of the terms which have not been 
        computed].
\end{enumerate} 
Finally, it would certainly be quite useful to compare the value of 
$a_x$ extracted from $K^+ \to 3 \pi$ decays vs.~the value 
extracted in a similar way from $K_L \to 3 \pi$ decays.

\section{Conclusions}
We have outlined a method that allows to systematically 
evaluate rescattering effects in $K \to 3 \pi$ decays 
by means of an expansion in powers of the 
$\pi\pi$ scattering lengths. 
This approach is less ambitious than the ordinary loop
expansion performed in effective field theories, 
such as CHPT: the scope is not a dynamical 
calculation of the entire decay amplitudes, 
but a systematical evaluation of the 
singular terms due to rescattering effects only. 
In particular, our main goal has been 
a systematical description of the cusp effect 
in \Kp00 \cite{Cabibbo:2004gq} in terms of the 
$\pi\pi$ scattering lengths. 
{}From this point of view, the approach we have 
proposed is more efficient and substantially 
simpler than the ordinary perturbative 
expansion of CHPT.

Using this method we have explicitly computed all the 
$O(a_i^2)$ corrections to the leading cusp effect in \Kp00, 
extending the results of Ref.~\cite{Cabibbo:2004gq}. 
As shown in figure \ref{fig:plot2}, these extra terms 
produce a small square-root behavior also above the \pppm 
singularity. 

The present work allows to reduce the theoretical error
on the extraction of  $a_0-a_2$ from an experimental analysis of the \Kp00 
spectrum to about 5\%. A similar level of theoretical accuracy 
is also achieved in the case of the $K_L \to 3\pi^0$ spectrum. 
This level of precision is probably not sufficient to 
fully exploit the potentially very accurate data of 
NA48, and is also quite above the error on the 
predictions of $a_0-a_2$ in Ref.~\cite{Colangelo}. To reach this 
level of precision, a complete evaluation of the 
$O(a_i^3)$ corrections and --- at the same time --- 
of the effects due to radiative corrections
is needed.

\section*{Acknowledgments}
We are grateful to Italo Mannelli, Luigi Di Lella, and other members of NA48 for 
discussions about the $K \to 3\pi$ analysis. We also thank
Gilberto Colangelo and Juerg Gasser for useful discussions
and comments on the manuscript.

\end{document}